\documentclass[aps,amsmath,preprint,epsf,superscriptaddress,nofootinbib,preprint]{revtex4-1}
\pdfoutput=1
\usepackage{graphicx}
\usepackage{bm}
\usepackage{color}
\usepackage{amsmath,amssymb,amsfonts}	
\usepackage{hyperref}
\usepackage{setspace}
\usepackage[samesize]{cancel}
\usepackage{tikz}

\def\slashchar#1{\setbox0=\hbox{$#1$} 
\dimen0=\wd0 
\setbox1=\hbox{/} \dimen1=\wd1 
\ifdim\dimen0>\dimen1 
\rlap{\hbox to \dimen0{\hfil/\hfil}} 
#1 
\else 
\rlap{\hbox to \dimen1{\hfil$#1$\hfil}} 
/ 
\fi}

\usepackage[utf8]{inputenc}
\usepackage{graphicx}
\usepackage{physics}
\usepackage{tikz}
\usepackage{slashed}
\usepackage{relsize}
\usepackage[all]{xy}
\usepackage{cancel}
\usepackage{verbatim}
\usepackage{ytableau}
\usepackage{youngtab}
\usepackage{young}

\usepackage{tabularx}
\newcolumntype{Y}{&gt;{\centering\arraybackslash}X}

\begin{document}

\baselineskip 0.7cm

\preprint{IPMU17-0087}
\bigskip

\title{Dynamical Clockwork Axions}
\author{Rupert Coy}
\email[e-mail: ]{rupert.coy@umontpellier.fr}
\affiliation{Laboratoire~C.~Coulomb~(L2C),~Univ.~Montpellier,~CNRS,~Montpellier,~France}
\author{Michele Frigerio}
\email[e-mail: ]{michele.frigerio@umontpellier.fr}
\affiliation{Laboratoire~C.~Coulomb~(L2C),~Univ.~Montpellier,~CNRS,~Montpellier,~France}
\author{Masahiro Ibe}
\email[e-mail: ]{ibe@icrr.u-tokyo.ac.jp}
\affiliation{Kavli IPMU (WPI), UTIAS, The University of Tokyo, Kashiwa, Chiba 277-8583, Japan}
\affiliation{ICRR, The University of Tokyo, Kashiwa, Chiba 277-8582, Japan}

\begin{abstract}
The clockwork mechanism is 
a novel method for generating a large separation between the dynamical scale and interaction scale of a theory. 
We demonstrate how the mechanism can arise from a sequence of strongly-coupled sectors.
This framework  avoids elementary scalar fields as well as \textit{ad hoc} continuous global symmetries,  both of which are subject
to serious stability issues. The clockwork factor, $q$, is determined by the consistency of the strong dynamics. 
The  preserved global $U(1)$ of the clockwork appears as an accidental symmetry, resulting from discrete or $U(1)$ gauge symmetries,
and it is spontaneously broken by the chiral condensates.
We apply such a dynamical clockwork to construct models with an effectively
invisible QCD axion from TeV-scale strong dynamics. The axion couplings are determined by the 
localisation of the Standard Model interactions along the clockwork sequence.
The TeV spectrum includes either coloured hadrons or vector-like quarks. Dark matter can be accounted
for by the axion or the lightest neutral baryons, which are accidentally stable.
\end{abstract}

\maketitle

\tableofcontents

\section{Introduction}

The principle of naturalness~\cite{tHooft:1979rat} has been one of the most important guidelines in searching for physics beyond the Standard Model (SM). 
According to this principle, small parameters are not expected in a theory unless the theory becomes more symmetric in the limit that those parameters vanish. 
By applying this principle to higher dimensional interaction terms 
in a low energy theory 
resulting from new dynamics, interaction scales are usually identified 
as the energy scales of the dynamics. 
The pion decay constant and the Fermi constant are good examples of where interaction scales are close to the energy scales of the underlying dynamics.

The clockwork mechanism in \cite{ChoiIm,KaplanRattazzi,GiudiceMcCullough} 
provides a novel method of circumventing the principle of naturalness
(see \cite{Kim:2004rp} for an earlier work in the context of natural inflation).
It allows a  large separation between the interaction scale and the dynamical scale in a theory
with only ${\cal O}(1)$ parameters.
Recently, this mechanism has been harnessed in models of inflation \cite{KehagiasRiotto}, the axion \cite{Farina,Higaki:2016yqk}, composite Higgs \cite{AhmedDillon},  WIMPs \cite{Hambye:2016qkf}, and the relaxion 
\cite{Graham:2015cka,DiChiara:2015euo,Evans:2016htp}.

The basic clockwork idea involves many sites of spontaneous $U(1)$ symmetry breaking at a scale $f$,
each of which is associated with a Nambu-Goldstone boson (NGB).
The $U(1)$ symmetries are also explicitly broken down to a single $U(1)$ symmetry, $U(1)_0$, by couplings between the sites. 
Accordingly, only one linear combination of the NGBs remains massless.
Remarkably, the effective decay constant of the remaining NGB, $F$, can be much larger than $f$ 
when the $U(1)_0$ charges are appropriately chosen.
In particular, the clockwork mechanism is achieved by a very efficient arrangement of the symmetry,
which leads to an exponentially enhanced effective decay constant, $F \sim q^N f$, 
where $q>1$ is the clockwork factor and $N$ the number of sectors.%
\footnote{Ref.~\cite{GiudiceMcCullough} also provides generalisations of the clockwork mechanism to the interactions of fermions, gauge bosons, and gravitons.
 The range of applicability of the clockwork idea is discussed in Refs.~\cite{Craig:2017cda,Giudice:2017suc}.}
We note that essentially the same idea of separating the interaction scale and dynamical scale 
by multiple $U(1)$ symmetry breaking was previously proposed in Refs.~\cite{Harigaya:2014eta,*Harigaya:2014rga},
under the name of the phase-locking mechanism.
In this mechanism, however, the charge assignment of the $U(1)$ symmetries was 
not specified, and hence no exponential hierarchy was discussed.

The authors of Refs.~\cite{KaplanRattazzi,GiudiceMcCullough} showed how the large number of approximate global $U(1)$ symmetries, as well as 
the $U(1)_0$ charges in geometric progression with ratio $q$, can be attributed to extra-dimensional setups.
In this paper, we discuss alternative routes to achieve these important features for the clockwork mechanism by utilising strong dynamics at each site.
In our models, the clockwork factor is solely determined by the consistency of the strong dynamics.
We also discuss models where the global $U(1)_0$ symmetry appears as an accidental 
symmetry, resulting from discrete symmetries or $U(1)$ gauge symmetries. 

As for the phenomenological applications of the dynamical clockwork, we will focus on the axion solution to the strong CP problem,
which requires spontaneous breaking of a Peccei-Quinn symmetry, $U(1)_{PQ}$, with a decay constant large enough to comply with experimental/astrophysical searches
of the axion, $f_{PQ}\gtrsim 4\times10^{8}$ GeV~\cite[for review]{Raffelt:2006cw,Agashe:2014kda}.%
\footnote{Here, we assume for definiteness the KSVZ axion model~\cite{Kim:1979if,Shifman:1979if}.}
 It has been known for a long time \cite{Kim:1984pt} that strong dynamics can realise this mechanism with
heavy fermions charged under QCD as well as under a new gauge interaction that confines at scale $\Lambda\gtrsim f_{PQ}$: the heavy-fermion
condensate breaks $U(1)_{PQ}$ spontaneously and yields a composite axion. By using the clockwork, we will show how $\Lambda$ can be lowered
to the multi-TeV scale while keeping the axion sufficiently `invisible'.%
\footnote{ An additional motivation for strong dynamics in the multi-TeV range is to solve the gauge hierarchy problem
by the compositeness of the Higgs boson. Indeed, the simplest ultraviolet-complete composite Higgs models (see e.g. Refs.~\cite{Barnard:2013zea,Ferretti:2013kya,Cacciapaglia:2014uja,
Vecchi:2015fma,Bizot:2016zyu}) have some similarities with composite axion models:
they require heavy fermions charged under the new confining interaction as well as  QCD 
to yield composite top-quark partners.}

It is notable that a low PQ symmetry breaking scale guarantees better protection of the axion potential against explicit breaking effects from the Planck scale, $M_P$,
that are suppressed by powers of $\Lambda/M_P$~\cite{Hawking:1987mz,Lavrelashvili:1987jg,Giddings:1988cx,Coleman:1988tj,Gilbert:1989nq}. 
In general, for $\Lambda$ in the multi-TeV scale, this suppression is close but not quite sufficient 
to solve the strong CP problem, therefore we will discuss additional ways to protect the axion potential, specific to our dynamical clockwork models.
Different ideas to screen quantum gravity corrections can be found in 
Refs.~\cite{Kim:1981bb,Georgi:1981pu,Dimopoulos:1982my,Frampton:1981qu,Kang:1982bx,Lazarides:1982tw,
Banks:1991mb,Barr:1992qq,Kamionkowski:1992mf,Holman:1992us,Dine:1992vx,Dias:2002gg,Dvali:2005an,
Carpenter:2009zs,Harigaya:2013vja,Harigaya:2015soa,Redi:2016esr,Fukuda:2017ylt}.

In section \ref{DynamicalClockwork}, we present models where strong dynamics provides both spontaneous symmetry breaking at each clockwork site
and the link between neighbouring sites.
In section \ref{modules}, we construct alternative models where each strong sector is confined to a single site and 
the connection between sites is enforced by other exact symmetries.
In section \ref{InvisibleAxion}, we discuss the coupling between our dynamical clockwork models and QCD, in order to realise
the axion solution to the strong CP problem.
Finally, in section \ref{phenocosmo}, we describe the phenomenological and cosmological implications of  the dynamical clockwork axion.


\section{Clockwork linked by strong dynamics}
\label{DynamicalClockwork}

The standard clockwork mechanism \cite{ChoiIm,KaplanRattazzi,GiudiceMcCullough} involves a large number of fields which obey a $U(1)^{N+1}$ global symmetry 
that is broken spontaneously.
The $U(1)^{N+1}$ global symmetry  is also broken explicitly to a single global symmetry, $U(1)_0$, by ``nearest neighbour" couplings between the fields. 
In the simplest case, these fields are $N+1$ complex scalars, and the global $U(1)^{N+1}$ symmetry is spontaneously broken at some scale, $f$, 
then explicitly broken down to the $U(1)_0$ at a much lower scale. 
The potential takes the form \cite{KaplanRattazzi}
\begin{equation}
V(\phi) = \sum \limits_{j=0}^N \left(-m^2|\phi_j|^2  + \frac{\lambda}{4} |\phi_j|^4 \right) - \sum \limits_{j=0}^{N-1} \left(\frac{\epsilon}{f^{q-3}} \phi_j^\dagger \phi_{j+1}^q + h.c. \right),
\label{ClockworkPotential}
\end{equation}
such that $\epsilon \ll \lambda$. 
It is assumed here that $f$ is the same for each $\phi_j$, and relaxing this assumption does not alter the implementation of the clockwork mechanism. 

Below the scale $f$, we can rewrite the potential \eqref{ClockworkPotential} 
in terms of the $N+1$ Goldstone boson fields, $\phi_j \rightarrow U_j \equiv f e^{i\pi_j/(\sqrt{2}f)}$, which are associated with the spontaneously broken symmetries. This gives
\begin{equation}
V(\pi)  = - 2\epsilon f^4 \sum \limits_{j=0}^{N-1} \cos\frac{\pi_j - q \pi_{j+1}}{\sqrt{2}f}\simeq {\rm const}~+  \frac{\epsilon f^2}{2} \sum \limits_{j=0}^{N-1} \left(\pi_j - q \pi_{j+1}\right)^2.
\label{potmin}\end{equation}
The theory contains one massless exact NGB, corresponding to the remnant global $U(1)_0$ symmetry. 
Rotating into the mass basis, $\pi_j = O_{jk} a_k$, where $O \in O(N+1)$, we denote the Goldstone boson as $a_0$ and consequently $O_{j0} \propto q^{-j}$. 
This is the crucial feature of the clockwork mechanism: the massless axion component of each $\pi_j$ shrinks exponentially with increasing $j$, for $q>1$. 

By coupling the SM to one end of this chain of fields  ($\phi_N$ in the case $q>1$, $\phi_0$ in the case $q<1$), this mechanism can create an effective coupling 
	much smaller than one ($q^{-N}$ for $q>1$, $q^N$ for $q<1$) from $\mathcal{O}(1)$ interactions in the Lagrangian. 
For $q \neq 1$ and large $N$, it is possible to generate very substantial differences in scale.

The notable features of the clockwork mechanism are a) the nearest neighbour nature of the interactions, and b) the large number of approximate global $U(1)$ symmetries. 
The nearest neighbour nature of the interactions is technically achieved by a special arrangement of the $U(1)_0$ charges in geometric progression with a ratio $q$.
To make the construction more convincing, however, some justification of the origin of such a well-organised charge assignment is required.  
Besides, given the arguments that all global symmetries are broken by quantum gravity effects~\cite{Hawking:1987mz,Lavrelashvili:1987jg,Giddings:1988cx,Coleman:1988tj,Gilbert:1989nq,Banks:2010zn}, models may seem more plausible if they do not rely on global symmetries.

In the following, we discuss how the well-organised $U(1)_0$ charge assignment is achieved by utilising strong dynamics. 
The spontaneous symmetry breaking is achieved by the confinement of asymptotically free gauge theories at each site.
In the first class of models, we find that the $U(1)_0$ charges in geometric progression are automatically obtained from the fermion content
of the chain of gauge theories.
We also show that such dynamical clockwork can be realised without imposing any continuous global symmetry.

\subsection{Dynamical phase locking}
\label{Model1}

First, we consider a model with an $SU(N_c)^N$ gauge symmetry containing $N+1$ vector-like fermions, $\psi_j$, $j=0,\ldots, N$, 
which transform as the fundamental, $\bf F$, of $SU(N_c)_{j+1}$ and as some representation, $\mathbf{R}$, of $SU(N_c)_{j}$ other than 
the fundamental representation. 
We note the ``boundary conditions" that $\psi_0$ and $\psi_N$ transform only under a single $SU(N_c)$. 
This is shown in Table \ref{ModelRepns1}. 
We assume that each $SU(N_c)_j$ is asymptotically free and exhibits confinement at the dynamical scale $\Lambda_j$.
We also assume that the fermion mass terms allowed by the gauge symmetries are all vanishing, which is natural in the 't Hooft sense 
(we will deal later with global symmetry protection against gravity).

Due to confinement, the axial symmetry, $U(1)^{(j)}_A$, associated with each $\psi_j$ is spontaneously broken by
\begin{eqnarray}
\langle{\overline{\psi}_i{\psi_j} }\rangle = \tilde{\Lambda}_j^3 \delta_{ij} \ ,
\label{condensate}
\end{eqnarray}
where we omit gauge indices and define $\tilde{\Lambda}_j \simeq \max\{\Lambda_j, \Lambda_{j+1}\}$, with  $\Lambda_0 = \Lambda_{N+1} = 0$.%
\footnote{More precisely, when the condensate forms at $\tilde\Lambda_j$, the naive-dimensional-analysis, large-$N_c$ estimate is
$\langle{(\overline{\psi_i})_a({\psi_j})^b }\rangle =  \delta_{ij}\delta_a^b \ \tilde N_c\tilde{\Lambda}_j^3 /(16\pi^2)$, where $\tilde N_c$ is the dimension
of $\psi_j$ w.r.t. the confining group, while $a,b$ are `flavour' indices of the other $SU(N_c)$ group, which confines at a smaller scale.
The associated NGB decay constant is $f_j = \sqrt{\tilde N_c} \tilde{\Lambda}_j /(4\pi)$.
In the following we will adopt the schematic notation of Eq.~(\ref{condensate}), which is sufficient for our purposes.}

The quantum anomaly of each $SU(N_c)_j$ gauge symmetry also breaks the axial symmetries $U(1)_A^{(j-1)}$ and $U(1)_A^{(j)}$ explicitly.
Thus, the Goldstone bosons, $\pi_{j}$, associated with the chiral symmetry breaking of $U(1)_A^{(j)}$, 
$\overline{\psi}_{Rj} \psi_{Lj} \sim \tilde{\Lambda}_j^3 e^{i\pi_j}$, obtain non-vanishing mass terms due to the anomalous breaking.%
\footnote{We denote the four-component Dirac fermions by $\psi$ and the left-handed two-component Weyl fermions by
$\psi_L$ and $\overline{\psi}_R$,  
with $ \psi = (\psi_L, \overline{\psi}_R^\dagger)^T$.
}
There exists, however, a linear combination of the $U(1)_A$ axial symmetries   which survives  the anomalous breaking of $SU(N_c)^N$.
Accordingly, there is one remaining massless NGB, the axion $a$.
Other pseudo-Goldstone modes, on the other hand, obtain masses at the dynamical scale $\Lambda$ (we assume for simplicity $\Lambda_k \sim \Lambda$ for all $k$), as in the case of the $\eta'$ meson in QCD.\footnote{Note that,in QCD with massless quarks and $N_f$ flavours, there are also exact NGBs in the adjoint of $SU(N_f)$.
Analogously, in the present model with $\psi_j\sim ({\bf R}_j,{\bf F}_{j+1})$, when $SU(N_c)_j$ [$SU(N_c)_{j+1}$] confines, one is left with NGBs in the adjoint of $SU(N_f)$, with $N_f = {\rm dim}({\bf F})$ [$N_f = {\rm dim}({\bf R})$]. However, one can check that these NGBs are all charged under the other gauge group,
which eventually confines as well, giving them a mass of order $\Lambda$. The only gauge singlets are associated with the $U(1)_A^{(j)}$ symmetries
as discussed above. 
}

\begin{table}[t]
\caption{\small\sl Structure of nearest neighbour couplings embedded in an $SU(N_c)^N$ gauge theory.
Fermions in each site are vector-like Dirac fermions.
For an application to the invisible QCD axion model in section \ref{InvisibleAxion},
we also show the modified boundary condition, where $\psi_N$ transforms under the QCD gauge group, $SU(3)_{\rm QCD}$. 
}
 \small
\begin{center}
\begin{tabular}{|c|c|c|c|c|c||c|} \hline
& $SU(N_c)_1$ & $SU(N_c)_2$ & $SU(N_c)_3$ & \ldots & $SU(N_c)_N$ & $SU(3)_{\rm QCD}$ \\ \hline
$\psi_0$ & $\mathbf{N_c}$ & $\mathbf{1}$ & $\mathbf{1}$ &\ldots & $\mathbf{1}$ &$\mathbf{1}$\\ \hline
$\psi_1$ & $\mathbf{R}$ & $\mathbf{N_c}$ & $\mathbf{1}$ &$\ddots$  & $\mathbf{1}$& $\mathbf{1}$\\ \hline
$\psi_2$ & $\mathbf{1}$ & $\mathbf{R}$ & $\mathbf{N_c}$ &$\ddots$ & $\mathbf{1}$ &$\mathbf{1}$\\ \hline
$\psi_3$ & $\mathbf{1}$ & $\mathbf{1}$ & $\mathbf{R}$ &$\ddots$ & $\mathbf{1}$ &$\mathbf{1}$\\ \hline
$\psi_4$ & $\mathbf{1}$ & $\mathbf{1}$ & $\mathbf{1}$ &$\ddots$ & $\mathbf{1}$ &$\mathbf{1}$\\ \hline
\vdots & \vdots & \vdots & \vdots & $\ddots$ & $\vdots$& $\vdots$\\ \hline
\vdots & \vdots & \vdots & \vdots & $\ddots$ & $\mathbf{N_c}$& $\vdots$\\ \hline
$\psi_N$ & $\mathbf{1}$ & $\mathbf{1}$ & $\mathbf{1}$ &\ldots & $\mathbf{R}$ &$\mathbf{3}$\\ \hline
\end{tabular}
\end{center}
\label{ModelRepns1}
\end{table}

Let us see more closely how the clockwork mechanism is realised. 
The anomaly-free $U(1)_A$ is given by a linear combination of the  axial currents,
\begin{eqnarray}
\label{jA}
j_{A}^\mu =  \sum_{j = 0}^{N}q_j \times j_j^\mu\ ,
\end{eqnarray}
where $j_j^\mu$ is defined by 
\begin{eqnarray}
j_j^\mu = \frac 12 \overline{\psi}_j \gamma_5 \gamma^\mu \psi_j\ .
\end{eqnarray}
The $q_j$ coefficients satisfy
\begin{eqnarray}
\label{anomalyfree}
\frac 12 \times q_0 + T({\mathbf R}) N_c \times q_1 &=& 0 \ , \\
\frac 12 d({\mathbf R})\times q_{j-1}+ T({\mathbf R}) N_c\times q_{j}  &=& 0 \ , \quad j=2,\dots,N-1 \ ,\\
\frac 12 d({\mathbf R})\times q_{N-1}+  T({\mathbf R}) \times q_{N}  &=& 0 \ , 
\label{anomalyfree3} 
\end{eqnarray}
where $T(\mathbf{R})$ and $d(\mathbf{R})$ are respectively the Dynkin index and the dimension of the representation $\mathbf{R}$,
and we used $T({\bf F})=1/2$ and $d({\bf F})=N_c$.
The solution to these conditions is given by
\begin{eqnarray}
\label{q0}
q_0 = 1\ , \quad 
q_j = \frac{1}{d({\mathbf R})} \times q^{-j} \quad (j=1,\dots,N-1) \ , \quad q_N = \frac{N_c}{d(\mathbf{R})} \times q^{-N} \ ,
\end{eqnarray}
where 
\begin{eqnarray}
q = - \frac{2T({\mathbf R})N_c}{d(\mathbf{R})} \ ,
\end{eqnarray}
up to an overall normalisation.
In this way, 
the $U(1)_A$ charge assignment in geometric progression with a ratio $q$
is obtained. 


The fermion bilinear terms contain the axion component as
\begin{eqnarray}
\label{bilinear}
\overline{\psi}_{Rj} \psi_{Lj} \sim \tilde{\Lambda}_j^3 \times e^{iq_j\frac{a}{f_a}}\ , 
\end{eqnarray}
where the axion field resides in the conserved current of Eq.\,\eqref{jA},
\begin{eqnarray}
j^\mu_A \sim f_a \partial^\mu a \ ,
\end{eqnarray}
with $f_a$ being the axion decay constant.
Given the current and charge normalisations in Eqs.\,\eqref{jA} and \eqref{q0}, 
for $q > 1$ the axion decay constant $f_a$ is determined by the dynamical scale of the first site,
therefore one expects $f_a \simeq f\sim \Lambda/4\pi$.%
\footnote{For $q<1$ instead, with the normalisation in Eq.\,\eqref{q0}, one expects $f_a \simeq q_N 
\times f$.} 
Thus, when the SM couples to the axion through the $N$-th site, the axion interactions are suppressed 
not by $f_a$ but by $F_a \simeq q^N f_a$.
Thus, our model provides a dynamical realisation of the clockwork mechanism,
where the phase rotations of fermion bilinears are locked by strong dynamics.

We have not yet specified the representation ${\mathbf R}$.
Since we assume asymptotically free gauge theories, we require that the one-loop $\beta$ functions are negative,
that is,
\begin{eqnarray}
\beta_j(g_j)  = -\frac{g_j^3}{(4\pi)^2} \left[
\frac{11}{3}N_c - \frac{2}{3}d({\mathbf R}) - \frac{4}{3} T({\mathbf R})N_c \right]  
< 0 
\ .
\end{eqnarray}
We also require that the clockwork factor is larger than $1$,
\begin{eqnarray}
2T({\mathbf R})N_c > d(\mathbf{R}) \ .
\end{eqnarray}
From these two conditions, we find that $\bf R$ should be the
two-index anti-symmetric representation, ${\mathbf A_2}$, with
\begin{eqnarray}
T({\mathbf A_2}) = \frac 12(N_c - 2) \ , \quad d({\mathbf A_2}) = \frac{1}{2}N_c (N_c -1)\ ,
\end{eqnarray}
and $N_c = 4 $ or $N_c  =5$.
 In these cases, we find that the clockwork factor is
 \begin{eqnarray}
q = - \frac{4}{3}\quad  (N_c = 4) \ ,  \quad q = - \frac{3}{2}\quad (N_c = 5) \ ,
\end{eqnarray}
which provide an exponential separation of scales
\begin{eqnarray}
10^{0.125\times N}\quad (N_c = 4)  \ , \quad 
10^{0.176\times N}\quad (N_c = 5)  \ .
\end{eqnarray}

 Indeed, in this class of models it is difficult to achieve a large clockwork factor, but one should remark than
even a mild separation of scales may have interesting phenomenological consequences for the axion. 
The smallness of $q$ is simply due to the group theoretical reasons above, and 
replacing $SU(N_c)$ with different gauge groups does not change the situation much. In section \ref{modules},
we will present a different realisation of the dynamical clockwork with a larger $q$, which is of order $N_c$ or even independent of the strong dynamics.

Before closing this section, let us rephrase the above arguments in terms of the effective Lagrangian of 
the NGB modes, $\pi_j$, associated with spontaneous $U(1)_A^{(j)}$ breaking.
Strictly speaking, the $\pi_j$'s are not well defined since the $U(1)_A^{(j)}$ symmetries are broken by anomalies of the strong dynamics.
Still, the effective Lagrangian provides us with a convenient description of the low energy theory below the dynamical scale,
as in the case of chiral perturbation theory with the $\eta'$.
The effective theory of $\pi_j$'s is given by
\begin{eqnarray}
\label{EFT}
{\cal L} = \frac{1}{2} \sum_{j=0}^{N} f_{j}^2 \partial \pi_j \partial \pi_j + F_{\rm anom}[ \pi_0,\ldots, \pi_N ] \ .
\end{eqnarray}
Here, we normalise $\pi_j$ to be dimensionless variables whose shifts correspond to the axial rotations of $\psi_j$. 
The decay constants are expected to be of order
$f_j \sim \tilde{\Lambda}_j /(4\pi)$.
The scalar potential, $F_{\rm anom}$, represents the explicit breaking of  $U(1)_A^{(j)}$ symmetries 
by the quantum anomalies, which depends periodically on $\pi_j$'s through the combinations
\begin{eqnarray}
&&\pi_0 + 2T({\mathbf R}) N_c\times \pi_1 \ , \\
&&d({\mathbf R})\times \pi_{j-1}+ 2T({\mathbf R}) N_c\times \pi_{j}\ ,\\
&&d({\mathbf R})\times \pi_{N-1}+  2T({\mathbf R})\times \pi_{N} \ , 
\end{eqnarray}
corresponding to the anomalies with respect to $SU(N_c)_1,\dots, SU(N_c)_N$, respectively.
As expected, the scalar potential has a flat direction, i.e. the axion direction,
\begin{eqnarray}
\label{pij}
\pi_j = q_j \frac{a}{f_a}+\dots \ ,
\end{eqnarray}
where $q_j$ is given in Eq.\,\eqref{q0}, is in agreement with Eq.\,\eqref{bilinear}.
Finally, by substituting Eq.\,\eqref{pij} into Eq.\,\eqref{EFT}, we also find that the 
axion decay constant is 
\begin{eqnarray}
f_a = \left( 
\sum_{j=0}^N q_j^2 f_j^2 
\right)^{1/2}  \ ,
\end{eqnarray}
for a canonically normalised axion.
When the dynamical scale in each site 
is approximately the same, i.e. $f_j \sim f$,
we find $f_a \sim f$. 

\subsection{Exact discrete symmetries}
\label{discretesymmetry}

So far, we have assumed the global $U(1)_A^{(j)}$ symmetries. These are explicitly broken by non-zero 
mass terms for the fermions, which could be induced in general by quantum gravity corrections.
By modifying the model slightly,  we can replace the $U(1)_A^{(j)}$ symmetries 
with discrete gauge symmetries, which are immune to quantum gravity 
effects~\cite{Krauss:1988zc,Preskill:1990bm,Preskill:1991kd,Banks:1991xj}. 
For this purpose, we change the representation of $\psi_0$ from ${\mathbf F}$
to some other representation ${\mathbf R_0}$ and allow $\psi_N$ to have $N_f$ flavours.%
\footnote{For $N_f > 1$, the chiral symmetry breaking by $SU(N_c)_N$ confinement leads 
to multiple NGB modes, in the adjoint of $SU(N_f)$. We will discuss this in section \ref{InvisibleAxion}.}
The anomaly-free conditions of $U(1)_A$ in Eqs.\,\eqref{anomalyfree} and \eqref{anomalyfree3} are also changed accordingly.

Now, we impose a ${\mathbb Z}_m^{N+1}$ discrete symmetry, where the left-handed components of the $\psi$'s 
transform as
\begin{eqnarray}
\psi_{jL} \to \psi_{jL}' = e^{i\frac{2\pi}{m}} \psi_{jL} \ , 
\end{eqnarray}
under ${\mathbb Z}_m^{(j)}$.
The  ${\mathbb Z}_m^{N+1}$ symmetry is free from $SU(N_c)^N$ anomalies when
\begin{eqnarray}
{\mathbb Z}_m^{(0)} : && \quad 2T({\mathbf R_0}) = 0 \quad ({\rm mod}\,\, m) \ , \\
{\mathbb Z}_m^{(j)} : && \quad N_c\times 2T({\mathbf R}) = 0\quad  \wedge 
\quad d({\mathbf R}) = 0 \quad ({\rm mod}\,\, m) \ , \\
{\mathbb Z}_m^{(N)} : && \quad N_f\times 2T({\mathbf R}) = 0 \quad ({\rm mod}\,\, m) \ , 
\end{eqnarray}
where ${\mathbf R} = {\mathbf A_2}$.
Thus, for $N_c = 4$, the model admits an anomaly-free ${\mathbb Z}_2^{N+1}$ symmetry 
when $2T({\mathbf R_0})$ is even.
For $N_c = 5$, on the other hand, the model admits an anomaly-free ${\mathbb Z}_5^{N+1}$ symmetry
when $2T({\mathbf R_0}) $ and $N_f$ are multiples of $5$.
Once we impose ${\mathbb Z}_m^{N+1}$ symmetries, the $U(1)_A$ symmetry appears as an approximate 
accidental symmetry.%
\footnote{The exact discrete symmetries forbid the fermion mass terms, but they may allow
higher-dimensional operators, suppressed by powers of $M_{P}$, that will break $U(1)_A^{(j)}$ at some level.}

The model with $N_c = 5$ is particularly interesting for ${\mathbf R_0} = \mathbf{Adj}$ and $N_f = 5$.%
\footnote{Here $\psi_0$ has independent left-handed and right-handed components, both in the adjoint representation.}
In this case, the one-loop beta functions of all the $SU(N_c)_{j = 1, \cdots , N}$ become the same,
with which all the confinement scales $\Lambda_k$ are approximately the same, s long as all the gauge coupling constants
are similar at the ultraviolet scale.
In section \ref{InvisibleAxion}, we will construct an invisible QCD axion model 
by identifying a subgroup of the flavour symmetry of $\psi_N$ with $SU(3)_{\rm QCD}$
without additional, unwanted NGB modes.

\section{Clockwork with connected strong-dynamics modules}
\label{modules}

In the class of models in subsection \ref{Model1}, the dynamics of every gauge group was intrinsically linked 
by a chain of fermions charged under both $SU(N_c)_j$ and $SU(N_c)_{j+1}$. 
This nice feature can perhaps be considered the gauge group equivalent 
to the nearest neighbour coupling of fields outlined in the original clockwork model. 
In this class of models, however, the clockwork factor is rather small due to the requirement of asymptotic freedom of the strong dynamics at each site.
In this section, we discuss alternative links between the sites which also permit the clockwork mechanism.
Each site is associated with a module of strong dynamics, 
and the connection is realised with the help of discrete `gauge' symmetries, or $U(1)$ gauge symmetries,
such that the axion $U(1)_A$ arises as a residual, accidental symmetry.

\subsection{Contact connection}
\label{Model2}

As a module of strong dynamics, we consider a model with an $SU(N_c)$ gauge symmetry containing two vector-like fermions, $Q$
and $\psi$, which transform as the representations ${\mathbf R}_Q$ and ${\mathbf R}_\psi$ of 
$SU(N_c)$, respectively. 
Several modules, numbered from $1$ to $N$, are illustrated in Table \ref{ModelRepns2}.
As in the previous section, we assume that each $SU(N_c)_j$ gauge theory is asymptotically free 
and exhibits confinement at the dynamical scale, $\Lambda_j$.

In each module, there is an anomaly-free axial current,
\begin{eqnarray}
\label{jAj}
j_{Aj}^\mu =  q_{Qj} \, j_{Qj}^\mu + q_{\psi j}\, j_{\psi j}^\mu  \qquad 
\left( j_{Qj}^\mu = \frac 12 \overline{Q}_j \gamma_5\gamma^\mu Q_j \ ,
\quad
j_{\psi j}^\mu = \frac 12 \overline{\psi}_j \gamma_5\gamma^\mu \psi_j  \right)\ ,
\end{eqnarray}
where the charges $q_Q$ and $q_\psi$ satisfy the anomaly-free condition
\begin{eqnarray}
\label{qANOM}
q_{Qj}T({\mathbf R}_Q) + q_{\psi j} T({\mathbf R_\psi}) = 0 \ .
\end{eqnarray}
Here the axial charges are defined by $q_f \equiv q(f_L) = -q(f_R)$, for $f=Q,\psi$.
Due to confinement, this anomaly-free axial $U(1)_A^{(j)}$ symmetry is broken spontaneously,
and we obtain a massless NGB in each module.

\begin{table}[t]
\caption{\sl \small The structure of connected modules of strong dynamics in the contact connection model.
The fermions, $Q$ and $\psi$, transform as the representations of ${\mathbf R}_Q$ and ${\mathbf R}_\psi$ of 
$SU(N_c)$, respectively.
The dashed lines show the links between modules via the interactions in Eq.\,\eqref{contact}.}
\begin{center}
\begin{tabular}{|c|c|} \hline
& $SU(N_c)_1$   \\ \hline
$Q_1$ & $\mathbf{R}_Q$  \\ \hline
$\psi_1$ & $\mathbf{R_\psi}$ \\ \hline
\end{tabular}
\hspace{-.3cm}
\raisebox{-1.em}{
\begin{tikzpicture}
\draw[densely dashed] (0,0) -- (0.4,0) -- (0.4,.5) -- (.8,.5);
\end{tikzpicture}
}
\hspace{-.35cm}
\begin{tabular}{|c|c|} \hline
& $SU(N_c)_2$   \\ \hline
$Q_2$ & $\mathbf{R}_Q$  \\ \hline
$\psi_2$ & $\mathbf{R_\psi}$ \\ \hline
\end{tabular}
\hspace{-.3cm}
\raisebox{-1.em}{
\begin{tikzpicture}
\draw[densely dashed] (0,0) -- (0.4,0) -- (0.4,.5) -- (.8,.5);
\end{tikzpicture}
}
\hspace{-.35cm}
\begin{tabular}{|c|c|} \hline
& $SU(N_c)_3$   \\ \hline
$Q_3$ & $\mathbf{R}_Q$  \\ \hline
$\psi_3$ & $\mathbf{R_\psi}$ \\ \hline
\end{tabular}
\hspace{-.3cm}
\raisebox{-1.em}{
\begin{tikzpicture}
\draw[densely dashed] (0,0) -- (0.4,0) -- (0.4,.5) -- (.8,.5);
\end{tikzpicture}
}
$\cdots$
\hspace{-.3cm}
\raisebox{-1.em}{
\begin{tikzpicture}
\draw[densely dashed] (0,0) -- (0.4,0) -- (0.4,.5) -- (.8,.5);
\end{tikzpicture}
}
\hspace{-.35cm}
\begin{tabular}{|c|c|} \hline
& $SU(N_c)_N$   \\ \hline
$Q_N$ & $\mathbf{R}_Q$  \\ \hline
$\psi_N$ & $\mathbf{R_\psi}$ \\ \hline
\end{tabular}
\label{ModelRepns2}
\end{center}
\end{table}

In order to connect the modules with each other, we introduce higher-dimensional contact-interaction terms,
\begin{eqnarray}
\label{contact}
{\cal L} = \sum_{j=1}^{N-1} \frac{1}{M_*^2} (\overline{\psi}_{Rj}{\psi_{jL}})^\dagger (\overline{Q}_{Rj+1} Q_{Lj+1}) +h.c.\ ,
\end{eqnarray}
where $M_*$ denotes a dimensionful parameter which is larger than the dynamical scale.
In the presence of the contact interactions, the $U(1)_A^{N}$ symmetries are broken down to a single $U(1)_A$ symmetry,
whose charge assignment satisfies
\begin{eqnarray}
\label{qMED}
q_{\psi j}  = q_{Q j+1} \ .
\end{eqnarray}
By solving Eqs.\,\eqref{qANOM} and \eqref{qMED}, we obtain the conserved $U(1)_A$ current,
\begin{eqnarray}
\label{jA2}
j_A^\mu = \sum_{j = 1}^{N} (q_{Qj} \, j_{Qj}^\mu + q_{\psi j}\, j_{\psi j}^\mu )\ ,
\end{eqnarray}
where the charges are in geometric progression,
\begin{eqnarray}
\label{q2}
q_{Q1}  &=& 1\ ,\quad q_{Q2} = q^{-1} \ , \quad \cdots \ , \quad q_{QN} = q^{-(N-1)} \ ,\\ 
q_{\psi 1}  &=& q^{-1}, \quad  q_{\psi 2} = q^{-2}
\ , \quad \cdots \ , \quad q_{\psi N} = q^{-N} \ ,
\end{eqnarray}
up to an overall normalisation, with the clockwork factor
\begin{eqnarray}
\label{CCq}
q = - \frac{T({\mathbf R_\psi})}{T({\mathbf R}_Q)} \ .
\end{eqnarray}

As in the models of the previous section, the fermion bilinear terms contain the axion component as
\begin{eqnarray}
\label{bilinear2}
\overline{Q}_{Rj} Q_{Lj} \sim {\Lambda}_j^3 \times e^{iq_{Q j}\frac{a}{f_a}}\ , \quad 
\overline{\psi}_{Rj} \psi_{Lj} \sim {\Lambda}_j^3 \times e^{iq_{\psi j}\frac{a}{f_a}}
\ , 
\end{eqnarray}
where the axion field resides in the conserved current in Eq.\,\eqref{jA2}, with
$j^\mu_A \sim f_a \partial^\mu a$,
$f_a$ being the axion decay constant.
Thus, again, when the SM couples to the axion through the $N$-th site, the SM-axion interactions are suppressed by $F_a \simeq q^{N} f_a$. 

We may rephrase the above clockwork mechanism in terms of the effective theory of the Goldstone modes,
\begin{eqnarray}
\label{EFT2}
{\cal L} \simeq \sum_{j=1}^{N} \left[\frac{1}{2} f_{j}^2 \partial \pi_j \partial \pi_j + \frac{1}{2}f_{j}^2 \partial \xi_j \partial \xi_j 
+ \kappa f_j^4(\pi_j - q \xi_j)^2 \right]
+\sum_{j=1}^{N-1} \kappa' \frac{f_j^3 f_{j+1}^3}{M_*^2} (\xi_j - \pi_{j+1} )^2 \ .
\end{eqnarray}
Here, the $\pi$'s and $\xi$'s are the dimensionless NGBs associated with the axial rotations of $Q$'s and $\psi$'s, respectively. 
The decay constants in each module are estimated to be  $f_i \sim \Lambda_i/4\pi$.
The $\kappa$-term represents the explicit breaking due to the $SU(N_c)$ anomaly, while the $\kappa'$-term describes 
the explicit breaking due to the contact interactions in Eq.\,\eqref{contact}.
The coefficients $\kappa$ and $\kappa'$ parametrise our inability to calculate the strong dynamics effects  
(one expects $\kappa\sim \kappa' \sim (4\pi)^2$, up to order one numbers).
Those terms give masses to $2N-1$ Goldstone modes, while leaving the axion massless.

From the effective Lagrangian, we find that the massless axion is distributed in the $2N$ Goldstone modes according to
\begin{eqnarray}
\pi_j =q_{Qj} \frac{a}{f_a} +\dots  \ ,  \quad
\xi_j =q_{\psi j} \frac{a}{f_a} +\dots \ ,  
\end{eqnarray}
where $q$'s are  given in Eq.\,\eqref{q2}.
By substituting these solutions in the NGB kinetic terms, we obtain the axion decay constant,
\begin{eqnarray}
f_a^2 = 
\sum_{j=1}^N (q_{Qj}^2 + q_{\psi j}^2) f_j^2 ~.
\end{eqnarray}
The other $2N-1$ pseudo-NGBs obtain non-zero masses:
$N$ pseudo-Goldstone modes with masses of ${\cal O}(\Lambda)$,
and $N-1$ pseudo-Goldstone modes with masses of ${\cal O}(\Lambda f/M_*)$.

Let us discuss how the nearest neighbour contact interactions in Eq.\,\eqref{contact} are organised by discrete symmetries.
For that purpose, we consider ${\mathbb Z}_{m}^{(Qj)}$ and ${\mathbb Z}_{m}^{(\psi j)}$ groups, 
under which the left-handed components of $Q_j$ and $\psi_j$ transform as
\begin{eqnarray}
Q_{jL} \to Q_{jL}' &=& e^{\frac{2\pi i }{m}} Q_{jL} \ , \\
\psi_{jL} \to \psi_{jL}' &=& e^{\frac{2\pi i }{m}} \psi_{jL} \ , 
\end{eqnarray}
respectively.
The anomaly-free conditions for these transformations are 
\begin{eqnarray}
{\mathbb Z}_m^{(Qj)} : && \quad 2T({\mathbf R}_Q) = 0 \quad ({\rm mod}\,\, m) \ , \\
{\mathbb Z}_m^{(\psi j)} : && \quad  2T({\mathbf R_\psi}) = 0 \quad ({\rm mod}\,\, m) \ , 
\end{eqnarray}
which can be satisfied by choosing ${\mathbf R}_Q$, ${\mathbf R_\psi}$ and $m$ appropriately.
Then the contact interactions are  restricted to the ones in Eq.\,\eqref{contact}
by imposing invariance with respect to the subgroup $ {\mathbb Z}_m^{N-1}\subset{\mathbb Z}_{m}^{(Q)N} \times{\mathbb Z}_m^{(\psi)N}$ symmetry,
defined by 
\begin{eqnarray}
\label{Zmj}
{\mathbb Z}_{m}^{(j)} : Q_{j+1L} \to Q_{j+1L}' = e^{\frac{2\pi i }{m}} Q_{j+1L} \ , \quad
\psi_{jL} \to \psi_{jL}' = e^{\frac{2\pi i }{m}} \psi_{jL}\ , \quad j = 1\cdots N-1\ . 
\end{eqnarray}
It should be noted that the axial $U(1)_A$ symmetry is now regarded as an accidental symmetry
once the exact ${\mathbb Z}_m^{N-1}$ symmetry is imposed.

As an example, let us consider the case with ${\mathbf R}_Q = {\mathbf A_2}$ and ${\mathbf R_\psi} = {\mathbf{Adj}}$:
\begin{eqnarray}
2T({\mathbf R}_Q ) &=& N_c - 2 \ , \qquad
2T({\mathbf R}_\psi ) = 2 N_c  \ .
\label{CCexample1a}
\end{eqnarray}
If, for instance, $N_c =4$, the ${\mathbb Z}_2^{N-1}$ symmetry is anomaly-free 
and hence it can be regarded as a gauged discrete symmetry.
Incidentally, this choice leads to a rather large clockwork factor,
\begin{eqnarray}
q = - \frac{2 N_c}{N_c - 2} =- 4 \ .
\label{CCexample1b}
\end{eqnarray}
As another example, we may consider ${\mathbf R}_Q = {\mathbf A_2}$ and ${\mathbf R_\psi} = {\mathbf{A_3}}$,
where 
\begin{eqnarray}
2T({\mathbf R}_Q ) &=& N_c - 2 \ , \qquad
2T({\mathbf R}_\psi ) = \frac{1}{2} (N_c-2)(N_c-3)  \ .
\label{CCexample2}
\end{eqnarray}
If, for instance, $N_c = 11$, the ${\mathbb Z}_9^{N-1}$ symmetry is anomaly-free, 
with again $q = - 4$.
In both examples one can easily check that the $SU(N_c)$ gauge theory is asymptotically free.

The contact interaction terms in Eq.\,\eqref{contact} can be replaced straightforwardly with 
heavy mediator interactions.
For that purpose, one can introduce $N-1$ complex scalar fields, $\phi_j$, with Yukawa couplings to the fermions,
\begin{equation}
\mathcal{L}_\phi = \sum_{j=1}^{N-1} y_j \phi_j\left(\overline{\psi}_{Rj} \psi_{Lj} + \overline{Q}_{Rj+1} Q_{Lj+1}\right) + h.c.~.
\label{MediatedInteractions}
\end{equation}
Here, we assume that $\phi_j$'s are appropriately charged under the ${\mathbb Z}_m^{N-1}$ symmetry in Eq.\,\eqref{Zmj}.
Then the contact interactions in Eq.\,\eqref{contact} are obtained by integrating out the scalar fields with masses of ${\cal O}(M_*)$.

\subsection{WiFi-connection}
\label{WiFiconnection}

We may consider axial $U(1)$ gauge interactions as another way to connect modules.
For this purpose, in each module there are two fermions, $Q_j$ and $\psi_j$, which are
in the fundamental representation of an $SU(N_c)_j$, 
as illustrated in Table \ref{WiFi}.
Thus, there are $2N$ axial phase rotations of the fermions, of which $N$ are broken by $SU(N_c)^N$ anomalies.
In each module, there is an anomaly-free axial current,
\begin{eqnarray}
j_A^{(j)\mu} = \frac{1}{2} \left(
\overline{Q}_j\gamma_5 \gamma^\mu Q_j
-  \overline{\psi}_j\gamma_5 \gamma^\mu \psi_j
\right)\ .
\end{eqnarray}
The associated $U(1)_A^{(j)}$ is also free from cubic and gravitational anomalies, as $Q_j$ and $\psi_j$ have opposite charges.

For reasons that will be clear in a moment, one needs an additional interaction to 
forbid bilinears $\overline{Q_R}\psi_L$ and $\overline{\psi_R}Q_L$.
A technical way to achieve this goal, without affecting the anomaly cancellation conditions, is to assign $Q_j$ and $\psi_j$ to the fundamental and anti-fundamental representation of an auxiliary symmetry $SU(N'_c)$
that also confines at some dynamical scale, $\Lambda'\sim \Lambda$.
An alternative way is to gauge a vector symmetry, $U(1)_V$, defined by the current $j_V^{\mu} = (
\overline{Q}\gamma^\mu Q - \overline{\psi} \gamma^\mu \psi )/2$.
Either way, one enforces that the vacuum direction satisfies $\langle \overline{Q}_R\psi_L\rangle =\langle \overline{\psi}_RQ_L\rangle = 0$.
These auxiliary symmetries are also shown in Table \ref{WiFi}.

To connect the modules with each other, we introduce $U(1)_g^{(j)} = U(1)_A^{(j)}-qU(1)_A^{(j+1)}$ gauge interactions,
where the charges are arranged so that the corresponding currents are
\begin{eqnarray}
 j^{(j)\mu} = j_A^{(j)\mu} - {q} j_A^{(j+1)\mu}\ .
\end{eqnarray}
Once the $U(1)_g^{(j=1\cdots N-1)}$ gauge symmetries are imposed, 
no fermion mass terms are allowed, and an additional global $U(1)_A$ symmetry appears as  
an accidental symmetry.
Due to the chiral condensation of $SU(N_c)_j$ in each module, 
the axial $U(1)_g^{(j=1\cdots N-1)}$ gauge symmetries
are spontaneously broken by $\langle\overline{Q}_RQ_L\rangle$ and $\langle\overline{\psi}_R\psi_L\rangle$.
The corresponding NGBs are absorbed by the gauge bosons via the Higgs mechanism. 
At the same time, the global $U(1)_A$ symmetry is also broken spontaneously, which results in a massless axion. 

\begin{table}[t]
\caption{\sl \small The structure of the modules of strong dynamics in the WiFi-connection model.
The wavy lines show the link between the modules via the axial gauge interactions $U(1)^{(j)}_A-qU(1)^{(j)}_{j+1}$.
Each module consists of two fermions, $Q_j$ and  $\psi_j$, transforming under a gauge interaction $SU(N_c)_j$.
We also show two possible auxiliary interactions, $SU(N'_c)_j$ and $U(1)_V^{(j)}$, which forbid $\overline{Q}_{j}\psi_{j}$ terms.}
\begin{center}
\hspace{-.2cm}
\begin{tabular}{|*{1}{>{\centering\arraybackslash}p{1.5cm}|}} \hline
\vspace{.2cm}
$\Large{\cal M}_1$ 
\vspace{.5cm}
\\ \hline
\end{tabular}
\hspace{-.2cm}
\raisebox{0em}{
\begin{tikzpicture}
\draw (0,0) sin (.05,.1) cos (.1,0) sin (.15,-.1) cos (.2,0) sin (.25,.1) cos (.3,0) sin (.35,-.1) cos (.4,0)
sin (.45,.1) cos (.5,0) sin (.55,-.1) cos (.6,0) sin (.65,.1) cos (.7,0) ;
\end{tikzpicture}
}
\hspace{-.25cm}
\begin{tabular}{|*{1}{>{\centering\arraybackslash}p{1.5cm}|}} \hline
\vspace{.2cm}
$\Large{\cal M}_2$ 
\vspace{.5cm}
\\ \hline
\end{tabular}
\hspace{-.2cm}
\raisebox{0em}{
\begin{tikzpicture}
\draw (0,0) sin (.05,.1) cos (.1,0) sin (.15,-.1) cos (.2,0) sin (.25,.1) cos (.3,0) sin (.35,-.1) cos (.4,0)
sin (.45,.1) cos (.5,0) sin (.55,-.1) cos (.6,0) sin (.65,.1) cos (.7,0) 
;
\end{tikzpicture}
}
\hspace{-.25cm}
\begin{tabular}{|*{1}{>{\centering\arraybackslash}p{1.5cm}|}} \hline
\vspace{.2cm}
$\Large{\cal M}_3$ 
\vspace{.5cm}
\\ \hline
\end{tabular}
\hspace{-.2cm}
\raisebox{0em}{
\begin{tikzpicture}
\draw (0,0) sin (.05,.1) cos (.1,0) sin (.15,-.1) cos (.2,0) sin (.25,.1) cos (.3,0) sin (.35,-.1) cos (.4,0)
sin (.45,.1) cos (.5,0) sin (.55,-.1) cos (.6,0) sin (.65,.1) cos (.7,0) 
;
\end{tikzpicture}
}
\hspace{-.25cm}
$\cdots$
\hspace{-.2cm}
\raisebox{0em}{
\begin{tikzpicture}
\draw (0,0) sin (.05,.1) cos (.1,0) sin (.15,-.1) cos (.2,0) sin (.25,.1) cos (.3,0) sin (.35,-.1) cos (.4,0)
sin (.45,.1) cos (.5,0) sin (.55,-.1) cos (.6,0) sin (.65,.1) cos (.7,0) 
;
\end{tikzpicture}
}
\hspace{-.25cm}
\begin{tabular}{|*{1}{>{\centering\arraybackslash}p{1.5cm}|}} \hline
\vspace{.2cm}
$\Large{\cal M}_N$ 
\vspace{.5cm}
\\ \hline
\end{tabular}
\\
\vspace{1cm}
\hspace{-.25cm}
\begin{tabular}{|*{1}{>{\centering\arraybackslash}p{1.5cm}|}} \hline
\vspace{.2cm}
$\Large{\cal M}_j$ 
\vspace{.5cm}
\\ \hline
\end{tabular}
\quad =\quad
\begin{tabular}{|c|c|c||c|c|} \hline
& $SU(N_c)_j$ & $U(1)_A^{(j)}$ & $SU(N'_c)_j$ & $U(1)_V^{(j)}$
\\ \hline
$Q_j$ & $\mathbf{N_c}$ & $1$ & $\mathbf{N'_c}$ & $1$
\\ \hline
$\psi_j$ & $\mathbf{N_c}$ & $-1$ & \raisebox{-.1em}{$\mathbf{\overline{N}_c'}$} &$-1$  
\\ \hline
\end{tabular}
\label{WiFi}
\end{center}
\end{table}

To find out the axion direction, it is particularly transparent to use 
the effective field theory of the Goldstone modes.
In the WiFi model, the effective Lagrangian is given by
\begin{eqnarray}
\label{EFT3}
{\cal L} &\simeq& \sum_{j=1}^{N}\frac{1}{2} f_{j}^2 
\left[\partial^\mu \pi_j + (q g_{j-1}A_{j-1}^\mu - g_jA_{j}^\mu) \right]^2 \nonumber\\
&+& \sum_{j=1}^{N}
\frac{1}{2}f_{j}^2 
\left[\partial^\mu \xi_j  - (q g_{j-1}A_{j-1}^\mu -g_j A_{j}^\mu)
\right]^2 
+ \sum_{j=1}^{N} \kappa f_j^4\left(\pi_j + \xi_j\right)^2 
\ .
\end{eqnarray}
Here, $A_{j = 1 \cdots N-1}^\mu$ denote the gauge fields of the $U(1)_g^{(j=1\cdots N-1)}$ symmetries, 
and $g_j$ denotes the gauge coupling constant.%
\footnote{We set $A_0^\mu = A_N^\mu = 0$ identically. }
The NGBs, $\pi_j$ and $\xi_j$, correspond to the axial components of
$Q_j$ and $\psi_j$, respectively.
The last term shows the effect of the $SU(N_c)_j$ anomalies, where 
$\kappa$ parametrises our inability to calculate the effects of strong dynamics.

From the effective Lagrangian, we find the axion components in each Goldstone mode,
\begin{eqnarray}
\label{WiFiaxion}
\pi_j &=& q_{Q_j} \frac{f_a^2}{f_j^2}\frac{a}{f_a} +\dots \ ,  
\\
\quad 
\xi_j &=& q_{\psi_j}\frac{f_a^2}{f_j^2} \frac{a}{f_a} +\dots \ , 
\end{eqnarray}
which are not absorbed by the Higgs mechanism.
The $U(1)_A$ charges, $q_{Q,q}$, are given by
\begin{eqnarray}
\label{q3}
q_{Q_1}  &=& 1\ ,\quad q_{Q_2} = q^{-1} \ ,\quad \cdots \ ,\quad q_{Q_N} = q^{-(N-1)} \ ,\\ 
q_{\psi_1}  &=& -1\ ,\quad q_{\psi_2} = - q^{-1} \ ,\quad \cdots \ ,\quad q_{\psi_N} = -q^{-(N-1)} \ .
\end{eqnarray}
We are guaranteed that $U(1)_A$ is free from all anomalies because each $U(1)^{(j)}_A$ is anomaly-free by construction.
For a canonically normalised axion, the decay constant is selected to be
\begin{eqnarray}
f_a = \left[ 
\sum_{j=1}^N (q_{Qj}^2 + q_{\psi j}^2) f_j^{-2} 
\right]^{-1/2}  \ ,
\end{eqnarray}
which is dominated by the highest charge contribution, assuming $f_j \simeq f$ for all sites.

As in the previous models, the axion interactions are suppressed not by $f_a$ but by $F_a \simeq q^N f_a$,
if the SM couples to the axion through the $N$-th site. 
It should be noted that the axion clockwork factor, $q$, is not given by a dynamical reason, rather 
by the choice of the charge ratio between the neighbouring sites.

We now confirm that the axion direction is gauge invariant.
Recalling that the Goldstone modes, $f_j \pi_j$ and $f_j \xi_j$,
have canonically normalised kinetic terms, the relation in Eq.\,\eqref{WiFiaxion} can be reverted to
\begin{eqnarray}
a &=& \sum_{j=1}^{N} \left(
q_{Qj} \frac{f_a}{f_j}\times (f_j \pi_j) 
+ q_{\psi j} \frac{f_a}{f_j} \times (f_j \xi_j)
\right) =
f_a\sum_{j=1}^{N} \left(q_{Qj} \pi_j + q_{\psi j}  \xi_j
\right)\ .
\end{eqnarray}
Thus, the axion is invariant under the gauge transformations
\begin{eqnarray}
\pi_j &\to& \pi_j + \alpha_j -q \alpha_{j-1} \ ,  \qquad
\xi_j \to  \xi_j - \alpha_j + q  \alpha_{j-1} \ ,
\end{eqnarray}
where $\alpha_{j}$ denotes the gauge transformation parameter of $U(1)_g^{(j)}$.

Finally, let us discuss how the global $U(1)_A$ symmetry can be broken explicitly. 
Higher dimensional terms localised at each site necessarily preserve it, because $U(1)_A$ charges 
are proportional to the gauge $U(1)_g^{(j)}$ charges in each module.
In fact, the global $U(1)_A$ is broken only by  ``non-local" terms such as
\begin{eqnarray}
{\cal L}_{\cancel{U(1)_A}} \propto \prod_{j = 1}^{N}(\overline{Q}_{Rj} Q_{Lj} )^{q^{N-j}}\ ,
\end{eqnarray}
which carry a $U(1)_A$ charge of order $q^{N-1}$.
Therefore the explicit breaking of the global $U(1)_A$ symmetry is highly suppressed,
guaranteeing in particular a strong protection against quantum gravity corrections.%
\footnote{Here, we assume $q$ is an integer for simplicity.}

\section{Clockwork extension of invisible axion models}
\label{InvisibleAxion}

The dynamical clockwork models discussed above may have different applications.
Here we discuss the possibility of implementing the axion solution of the strong-$CP$ problem~\cite{Peccei:1977hh,Peccei:1977ur,Weinberg:1977ma,Wilczek:1977pj}.
For successful models, the axion coupling to QCD should be suppressed by 
a large decay constant, $F_a \gtrsim 4\times 10^8$\,GeV, to evade a number of constraints imposed by extensive axion 
searches~\cite[for review]{Raffelt:2006cw,Agashe:2014kda}.
Conventionally, the large decay constant is tied to the scale of the spontaneous symmetry breaking scale
of the Peccei-Quinn (PQ) symmetry, $U(1)_{PQ}$.
In the clockwork mechanism (and more generally in the phase-locking mechanism), however,
the decay constant can be hierarchically different from the scale of the actual dynamics.
In the following, we discuss dynamical models where $F_a \gtrsim 4\times 10^{8}\,$GeV 
is obtained as an effectively enhanced decay constant, $F_a \sim q^{N} \times f$,
while the actual dynamics occurs at the much lower scale, $f$.

For our discussion, it is useful to recall the main features of the KSVZ invisible axion model~\cite{Kim:1979if,Shifman:1979if}.
In the simplest KSVZ model, the PQ symmetry is spontaneously broken by the VEV of a complex scalar, $\phi$,
whose phase plays the role of the QCD axion, 
\begin{eqnarray}
\phi \sim \frac{F_a}{\sqrt{2}} e^{i a/F_a}\ ,
\end{eqnarray}
with the PQ charge of $\phi$ being $1$.
The complex scalar couples to $N_f$ vector-like quarks in the fundamental representation of $SU(3)_{\rm QCD}$,
\begin{eqnarray}
{\cal L} = y \phi \overline{Q}_RQ_L + h.c. \ .
\end{eqnarray}
The PQ symmetry is identified as the axial $U(1)$ symmetry of the $Q$'s.
Below the scale of spontaneous breaking of the PQ symmetry, 
the axion couples to QCD as
\begin{eqnarray}
{\cal L} = \frac{g_s^2}{32\pi^2} 
\frac{N_f\, a}{F_a}
G\tilde{G} \ ,
\label{QCDanom}\end{eqnarray}
due to the $U(1)_{PQ}$ anomaly with respect to QCD.
Here, $g_s$ denotes the QCD gauge coupling and 
$G$ is the QCD field strength, whose Lorentz and colour indices are understood. 
As long as the PQ-symmetry is not broken by other sources than the QCD anomaly, the strong $CP$ problem is 
successfully solved. In particular, the vector-like quarks, $Q$, should have no bare mass.

\begin{table}[t]
\caption{\sl \small The charge assignment of the vector-like fermions, $Q$ and $S$, 
in the composite axion model.}
\begin{center}
\begin{tabular}{|c|c|c|} \hline
& $SU(N_c)_N$ & $SU(3)_{\rm QCD}$  \\ \hline
$Q$ & ${\mathbf{N_c}}$ & ${\mathbf 3}$\\ \hline
$S$ & ${\mathbf{N_c}}$ & ${\mathbf 1}$\\ \hline
\end{tabular}
\label{composite}
\end{center}
\end{table}

The  KSVZ axion model can also be realised as a composite Goldstone mode~\cite{Kim:1984pt}. 
To this end, an $SU(N_c)$  gauge theory is introduced with vector-like fermions charged under $SU(N_c)\times 
SU(3)_{\rm QCD}$ according to Table \ref{composite}.
This model possesses an axial $U(1)$ symmetry with charges
\begin{eqnarray}
\label{eq:SUNaxial}
q_Q=1\ ,\quad q_S=-3\ ,
\end{eqnarray}
which is free from the $SU(N_c)$ anomaly but broken by the QCD anomaly.%
\footnote{The corresponding axial currents are normalised as $\overline{Q}\gamma_5\gamma^\mu Q/2$
and $\overline{S}\gamma_5\gamma^\mu S/2$, respectively, so that Eq.~(\ref{QCDanom}) holds with $N_f=N_c$.}
Therefore, this symmetry is a PQ symmetry,
which is spontaneously broken due to the chiral condensation of $SU(N_c)$ at scale $F_a$, 
where the axion appears as a composite NGB.%
\footnote{As there are 4 flavours charged under $SU(N_c)$, $S$ and the three colours of $Q$,
the chiral symmetry breaking of the axial $SU(4)$ flavour symmetry leads to 15 pseudo-Goldstone
modes. The QCD-singlet one is the axion, while the $14$ additional modes become massive 
due to the explicit breaking of the axial $SU(4)$ by the QCD gauge interactions.}

As we will see shortly, these KSVZ axion models are 
implemented in the dynamical clockwork models straightforwardly 
by identifying the unbroken $U(1)_A$ symmetry of the previous sections
with the PQ symmetry. 
By appropriately introducing fields charged under QCD, $U(1)_A$ is broken only by the QCD anomaly 
and the associated pseudo-NGB plays the role of the QCD axion.
The effective coupling of the axion to QCD is suppressed by $F_a \sim q^N \times f$
if the QCD charged states couple to the axion through the $N$-th site.

\subsection{Invisible QCD axion in the dynamical phase-locking model}
\label{DPaxion}
In the clockwork model of section \ref{DynamicalClockwork} linked by strong dynamics,
the massless axion appears as the NGB associated to the $U(1)_A$ current in Eq.~\eqref{jA}.
To couple the axion to QCD, we slightly modify the $N$-th site, so that 
$\psi_N$ is charged under  $SU(3)_{\rm QCD}$ as the fundamental representation. This has been already indicated in Table \ref{ModelRepns1}.

Consequently, $U(1)_A$ is broken by the QCD anomaly,
\begin{eqnarray}
{\cal L} = \frac{g_s^2}{32\pi^2} 
{d({\mathbf R})q_N} \frac{ a}{f_a}
G\tilde{G} \ .
\label{linkQCD}\end{eqnarray}
Here, $q_N$ denotes the $U(1)_A$ charge of the $N$-th site, which is given by
\begin{eqnarray}
q_N = \frac{1}{d({\mathbf R})}\frac{N_c}{3} q^{-N}\ ,
\end{eqnarray}
which is slightly modified from the one in Eq.\,\eqref{q0}.
As a result, the effective coupling of the axion to QCD is suppressed by an enhanced decay constant,
\begin{eqnarray}
\label{DynamicalFa}
 F_a = \frac{3}{N_c} q^{N} \times f_a \gg f_N\ ,
\end{eqnarray}
for $N$ large.

The model predicts a QCD octet NGB, 
associated with the chiral symmetry breaking of the axial $SU(3)$ flavour symmetry
of $\psi_N$, that obtains a mass from QCD loops,
\begin{eqnarray}
\label{octet}
m_8^2 \simeq \frac{3C_2}{4\pi}\alpha_{s} \Lambda_{N}^2 \ ,
\end{eqnarray}
where $\alpha_s = g_s^2/4\pi$ and the quadratic Casimir for the adjoint representation is $C_2 = 3$.
The model also predicts baryonic states charged 
under QCD, such as
\begin{eqnarray}
 B \sim\underbrace{\psi_N\psi_N\cdots \psi_N}_{N_c}
   \ ,\label{locked_baryons}
\end{eqnarray}
whose mass is expected to be ${\cal O}(N_c\Lambda_N)$.%
\footnote{Here, we suppressed the Lorentz and gauge indices. For example, when $\psi_N$ transforms in
the $SU(N_c)_N$ representation, ${\bf R}={\bf A_2}$, the gauge indexes are contracted with two $\epsilon$-tensors.}
 Additional coloured hadrons including $\psi_{1,\cdots ,N-1}$ are also possible
 (note that $\psi_j$-number is a conserved quantity, for each $j$).

Let us comment on the explicit breaking of the PQ symmetry.
In general composite axion models, the $U(1)_A$ symmetry, and hence
the PQ symmetry, can be explicitly broken by the vector-like fermion mass terms. 
This easily spoils the PQ mechanism.
As we discussed in section \ref{discretesymmetry}, however,
the model allows discrete gauge symmetries which forbid the mass terms,
so that the PQ symmetry appears as an accidental symmetry,
at least at the renormalisable level.

In the $N_c=4$ model of section \ref{discretesymmetry}, the lowest dimensional operators which are allowed by the ${\mathbb Z}_2^{N+1}$
symmetry are 
\begin{eqnarray}
\label{PQB2}
{\cal L}_{\cancel{PQ}} \sim \frac{1}{M_{P}^{2}} (\overline{\psi}_{Ri}\psi_{Li})^2  + h.c. \ ,
\end{eqnarray}
with $M_{P} \simeq 2.4\times 10^{18}$\,GeV.
Such higher dimensional terms lead to additional terms in the axion potential,
\begin{eqnarray}
V_{\cancel{PQ}}  \sim
\frac{ f_i^4\Lambda_i^{2}}{M_P^{2}} \cos\left({ \pi_i } + \delta_{\cancel{PQ}i}\right) \ , 
\end{eqnarray}
where $\delta_{\cancel{PQ}i}$ denotes the phase of the coefficient of the term in Eq.~\eqref{PQB2}.
According to section \ref{DynamicalClockwork}, $\pi_i = q_i a/f + \dots$ with $q_i\sim q^{-i}$, therefore
 the most relevant correction to the axion potential comes from the $0$-th site, and we will implicitly take $i=0$ below. 
This correction should be added to the QCD-anomaly contribution, $V_{PQ} \sim m_a^2F_a^2 \cos(a/F_a)$.
As a result, the effective QCD theta angle at the minimum of the axion potential is shifted,
\begin{eqnarray}
{\mit \Delta}\theta_{\rm eff}  \equiv \frac{1}{F_a}{\mit \Delta} a
&\sim&  \frac{1}{F_a} \frac{1}{m_a^2} \frac{f^4\Lambda^{2}}{M_P^{2}} \frac{\delta_{\cancel{PQ}}}{f} \nonumber\\
&\sim& 10^{-8} \delta_{\cancel{PQ}} 
\left( \frac{\Lambda}{10^{3}\,\rm GeV}\right)^2
\left( \frac{f}{10^{3}\,\rm GeV}\right)^3
\left( \frac{F_a}{10^{9}\,\rm GeV}\right)
\ .
\end{eqnarray}
where we used 
\begin{eqnarray}
m_a \simeq \frac{\sqrt{z}}{1+z}\frac{f_\pi m_\pi}{F_a}\ ,
\end{eqnarray}
with $z\simeq 0.6$ denoting the ratio of the up and down quark masses, and $f_\pi\simeq 93$\,MeV 
and $m_\pi \simeq 135$\,MeV the decay constant and the mass of the neutral pion, respectively.
This result shows that the low scale dynamics of the clockwork axion guarantees a better protection 
of the axion potential against the quantum gravity effects with respect to ordinary invisible axion models.
For a dynamical scale as low as ${\cal O}(1)$\,TeV, one needs only a mild suppression of the operators in Eq.~\eqref{PQB2} 
to satisfy the current limit, 
$\theta < 10^{-10}$~\cite{Baker:2006ts}.

Similarly, in the $N_c=5$ model of section \ref{discretesymmetry}, the lowest dimensional operators which are allowed by the ${\mathbb Z}_5^{N+1}$
symmetry are 
\begin{eqnarray}
\label{PQB}
{\cal L}_{\cancel{PQ}} \sim \frac{1}{M_{P}^{11}} (\overline{\psi}_{Ri}\psi_{Li})^5  + h.c. \ .
\end{eqnarray}
In this case, the effective theta angle of QCD at the minimum of the axion potential is shifted by
\begin{eqnarray}
{\mit \Delta}\theta_{\rm eff}  \equiv \frac{1}{F_a}{\mit \Delta} a \sim
 \frac{1}{F_a} \frac{1}{m_a^2} \frac{f^{10}\Lambda^{5}}{M_P^{11}} \frac{\delta_{\cancel{PQ}}}{f}  ~,
\end{eqnarray}
which is highly suppressed, allowing for a dynamical scale as large as  $\Lambda \sim {\cal O}(10^{12})$\,GeV.

A caveat is that the $SU(3)_{\rm QCD}$ symmetry is identified with a subgroup of the flavour symmetry 
of $\psi_N$, e.g. $SU(3)_{\rm QCD} \subset SU(N_f=5)$ for the model with the  ${\mathbb Z}_5^{N+1}$ symmetry.
In this case, there are $24$  Goldstone modes in addition to the axion.
Among them, one colour octet and four colour triplets become massive due to QCD radiative corrections.
The remaining four colour-singlet Goldstone modes can be lifted by also gauging the $SU(N_f-3) = SU(2)$ 
subgroup of the flavour symmetry, with a dynamical scale much larger than QCD. Then an $SU(2)$ triplet NGB receives a mass
from $SU(2)$ radiative corrections and the remaining singlet NGB receives a mass from the $SU(2)$ anomaly.%
\footnote{It is tempting to embed the minimal gauge symmetry of the grand unified theory, $SU(5)_{\rm GUT}$,
into the $SU(5)$ flavour symmetry of $\psi_N$.
In this case, however, there appears another axion mode from the $SU(N_c)_N$ sector:  one of the 24 NGBs is a singlet under $SU(3)_{QCD}\times SU(2)_w$,
it receives only a $GUT$-suppressed mass, and it couples to the QCD anomaly with a decay constant $f$.
Therefore, the QCD axion is dominated by this mode, rather than the one with the enhanced decay constant $q^N f$.
}

We note in passing that the dynamical clockwork model is different in many respects to the 
composite accidental axion model~\cite{Redi:2016esr} based on the moose theory~\cite{Georgi:1985hf}.
In the latter model, the fermions are chiral and QCD couples to both ends of the chain:
these two features achieve the PQ symmetry
as an accidental symmetry. 
Instead, in the dynamical clockwork, the fermions in each site are vector-like and QCD couples to one end of the chain of sites only.
In this case, the accidental PQ symmetry is a consequence of the external ${\mathbb Z}_m^{N+1}$ symmetry.
At the same time, the above features prevent the composite accidental axion model from achieving the clockwork mechanism.
The continuum limits (i.e. $N\to \infty$) of these models are also different.
As discussed in Refs.~\cite{GiudiceMcCullough,Giudice:2017suc}, the continuum limit of the clockwork 
mechanism corresponds to a model in $5$-dimensional spacetime
where the zero-mode axion is localised at one endpoint of the extra dimension while
QCD is attached to the other end.
In the composite accidental axion model, the continuum limit corresponds to 
a $5$-dimensional model where the zero mode axion has a flat configuration in the extra dimension, and QCD also propagates in this extra dimension.

\subsection{Invisible QCD axion in the contact-connection model}
\label{CCaxion}

The invisible QCD axion can be implemented in the clockwork model of section \ref{Model2} by identifying
the $U(1)_A$ symmetry in Eq.\,\eqref{jA2} as the PQ symmetry.
A simple way to break $U(1)_A$ by the QCD anomaly is 
to introduce a $(N+1)$-th module, comprising of $N_f$ flavours of vector-like fermions, $Q_{N+1}$, 
in the fundamental representation of $SU(3)_{\rm QCD}$, as shown in Table \ref{axion2}.
This last module interacts with $\psi_{N}$ via a contact interaction, as in Eq.\,\eqref{contact},
that implies $q_{Q_{N+1}} = q_{\psi N} = q^{-N}$.

Since $Q_{N+1}$ does not participate in strong dynamics other than QCD, 
this model corresponds to the original KSVZ axion model, with the scalar $\phi$ replaced
by $\langle\overline{\psi}_N\psi_N\rangle$. 
The  axion coupling to  QCD  is
\begin{eqnarray}
{\cal L} = \frac{g_s^2}{32\pi^2} 
{N_f q_{Q_{N+1}}} \frac{ a}{f_a}
G\tilde{G} \ ,
\label{ContactQCDaxion}
\end{eqnarray}
therefore the effective axion decay constant is enhanced by a factor $q^{N}$ with respect to $f_a\simeq f$.
The quality of the PQ symmetry can be guaranteed by imposing an exact ${\mathbb Z}_m^{(j=1\cdots N)}$ symmetry,
which requires $N_f=m$, as discussed in section \ref{Model2}.
Since $Q_{N+1}$ does not participate to the strong dynamics, the model also predicts vector-like coloured fermions with masses 
${\cal O}(\Lambda_N^3/M_*^2)$,  therefore $\Lambda_N$ should be well above the TeV scale if $\Lambda_N/M_*\ll 1$. 

\begin{table}[t]
\caption{\sl \small The termination of the chain of connected modules, which breaks the 
$U(1)_A$ symmetry by the QCD anomaly.
The model can be made consistent with a ${\mathbb Z}_m^{(j=1\cdots N)}$ symmetry, defined in section \ref{Model2},
by introducing $N_f = m$ flavours of $Q_{N+1}$.
}
\begin{center}
$\cdots$
\raisebox{-1.em}{
\begin{tikzpicture}
\draw[densely dashed] (0,0) -- (0.4,0) -- (0.4,.5) -- (.8,.5);
\end{tikzpicture}
}
\hspace{-.35cm}
\begin{tabular}{|c|c|} \hline
& $SU(N_c)_N$   \\ \hline
$Q_N$ & $\mathbf{R}_Q$  \\ \hline
$\psi_N$ & $\mathbf{R_\psi}$ \\ \hline
\end{tabular}
\hspace{-.3cm}
\raisebox{-1.em}{
\begin{tikzpicture}
\draw[densely dashed] (0,0) -- (0.4,0) -- (0.4,.3) -- (.8,.3);
\end{tikzpicture}
}
\hspace{-.35cm}
\begin{tabular}{|c|c|} \hline
& $SU(3)_{\rm QCD}$   \\ \hline
$Q_{N+1}$ & $\mathbf{3}$  \\ \hline
\end{tabular}
\label{axion2}
\end{center}
\end{table}
\begin{table}[t]
\caption{\sl \small
Another possible termination of the chain of connected modules, which breaks  the $U(1)_A$ symmetry by the QCD anomaly.
}
\begin{center}
$\cdots$
\raisebox{-1.em}{
\begin{tikzpicture}
\draw[densely dashed] (0,0) -- (0.4,0) -- (0.4,.5) -- (.8,.5);
\end{tikzpicture}
}
\hspace{-.35cm}
\begin{tabular}{|c|c|c|} \hline
& $SU(N_c)_N$  & $SU(3)_{\rm QCD}$ \\ \hline
$Q_N$ & $\mathbf{R}_Q$ & ${\mathbf 3}$\\ \hline
$\psi_N$ & $\mathbf{R_\psi}$ & $\mathbf 1$\\ \hline
\end{tabular}
\label{axionComp}
\end{center}
\end{table}

There is another simple way to couple the contact-connection clockwork axion to the QCD anomaly
that mimics the original composite axion model.
It consists of modifying the $N$-th module by charging $Q_N$ under QCD, as shown in
Table \ref{axionComp}, which indeed has the same structure as in Table \ref{composite}.
In this case the anomalous coupling of the axion to the QCD is given by
\begin{eqnarray}
{\cal L} = \frac{g_s^2}{32\pi^2} 
{d({\mathbf R_Q})q_{Q_N}} \frac{ a}{f_a}
G\tilde{G} \ ,
\label{aGG2bis}
\end{eqnarray}
with 
$q_{Q_N} = q^{-(N-1)}$,
where $q$ is given in Eq.\,\eqref{CCq}. Indeed, this realisation resembles closely the one of section \ref{DPaxion}
(compare with Eq.~\eqref{linkQCD}) and $Q_N$'s form coloured hadrons with mass ${\cal O}(\Lambda_N)$ or larger.

\subsection{Invisible QCD axion in the WiFi-connection model?}

In the WiFi-connection model of  section \ref{WiFiconnection},
the global $U(1)_A$ symmetry cannot be broken by effects localised in a single module.
This is because the $U(1)_A$ charge assignment is the same as for the 
gauged symmetry, $U(1)_g^{j=1,\cdots ,N-1}$, at each site $j$. Therefore no gauge invariant
term localised in one site can break the $U(1)_A$ symmetry.
This implies that the unbroken $U(1)_A$ cannot be identified with $U(1)_{PQ}$ if QCD is coupled to just one site.

To explicitly illustrate the impossibility of breaking the $U(1)_A$ symmetry, we introduce a $(N+1)$-th module 
as in Table \ref{axion3}, connected to the clockwork chain by a gauged $U(1)_g^{(N)}$ symmetry.
Once this module is introduced, the axial rotations are broken by the QCD anomaly,
\begin{eqnarray}
{\cal L} &=& \frac{g_s^2}{32\pi^2} 
\left({\mathbf N_c}\pi_{N+1}
+
{\mathbf N_c}\xi_{N+1}
\right)
G\tilde{G} \ .
\end{eqnarray}
The effective Lagrangian of the NGB modes contains
\begin{eqnarray}
\label{EFT4}
{\cal L} &\simeq& -  g_N f_a (q_{Q_N} - q_{\psi_N})\partial_\mu a A_{N}^\mu
+\frac{1}{2} f_{N+1}^2 
\left(\partial^\mu \pi_{N+1} +q  g_N A_{N}^\mu \right)^2 \nonumber\\
&+&
\frac{1}{2}f_{N+1}^2 
\left(\partial^\mu \xi_{N+1} - q  g_N  A_{N}^\mu
\right)^2 + \kappa f_{N+1}^4\left(\pi_{N+1} + \xi_{N+1}\right)^2 
\ ,
\end{eqnarray}
where the first term comes from the $N$-th site contributions after substituting the axion components 
of $\pi_N$ and $\xi_N$ in Eq.\,\eqref{WiFiaxion}.
Thus, $U(1)_g^{(N)}$ gauge invariance implies that the axion components in $\pi_{N+1}$ and $\xi_{N+1}$ are 
\begin{eqnarray}
\pi_{N+1} =  q^{-1} q_{Q_N} \frac{f_a^2}{f_{N+1}^2} \frac{a}{f_a}+\dots \ , \quad
\xi_{N+1} =  q^{-1} q_{\psi_N} \frac{f_a^2}{f_{N+1}^2} \frac{a}{f_a}+\dots \ .
\end{eqnarray}
Since $q_{Q_N}=-q_{\psi_N}$, the axion does not couple to the QCD anomaly.

We conclude that in the WiFi-connection model it is not possible to couple the $U(1)_A$ axion to QCD through the last site of the clockwork.
Nonetheless, independently from QCD, the WiFi models provides a clockwork axion mode that is exponentially localised with decay constants decreasing from $f_a$ in the first site to
$F_a \sim q^N \times f_a$ in the last one. Due to the gauge protection, this axion potential is extremely flat, which may have different applications.

\begin{table}[t]
\caption{\sl \small The termination of the chain of WiFi-connected modules.}
\begin{center}
$\cdots$
\hspace{-.2cm}
\raisebox{0em}{
\begin{tikzpicture}
\draw (0,0) sin (.05,.1) cos (.1,0) sin (.15,-.1) cos (.2,0) sin (.25,.1) cos (.3,0) sin (.35,-.1) cos (.4,0)
sin (.45,.1) cos (.5,0) sin (.55,-.1) cos (.6,0) sin (.65,.1) cos (.7,0) ;
\end{tikzpicture}
}
\begin{tabular}{|*{1}{>{\centering\arraybackslash}p{1.5cm}|}} \hline
\vspace{.2cm}
$\Large{\cal M}_{N+1}$ 
\vspace{.5cm}
\\ \hline
\end{tabular}
\quad =\quad
\begin{tabular}{|c|c|c||c|} \hline
& $SU(N_c)_{N+1}$ & $U(1)_A^{(N+1)} $ &  $SU(3)_{\rm QCD}$ 
\\ \hline
$Q_{N+1}$ & $\mathbf{N_c}$ & $1$ & $\mathbf{3}$ 
\\ \hline
$\psi_{N+1}$ & $\mathbf{N_c}$ & $-1$ & \raisebox{-0.1em}{$\overline{\mathbf{3}}$}   
\\ \hline
\end{tabular}\label{axion3}
\end{center}
\end{table}

\section{Phenomenological and cosmological implications}
\label{phenocosmo}

As we have discussed above, the dynamical clockwork mechanism can be 
used to achieve invisible QCD axion models, for which the effective decay constant 
is $F_a >  4 \times 10^{8}$\,GeV \cite{Raffelt:2006cw,Agashe:2014kda}.
Due to the possibly large separation between the dynamical scale and $F_a$,
it is obviously interesting to ask whether the models are directly testable at 
collider experiments by taking the dynamical scale to be $\mathcal{O}(1)$\,TeV. 
 We first outline the relevant collider searches, before describing the cosmological features of the models.

\subsection{Collider phenomenology}
\label{collider}

In the dynamical phase-locking model discussed in section \ref{DPaxion}, 
the extra quarks required for the QCD axion, which are charged under  $SU(3)_{\rm QCD}$,
are also charged under the new strong dynamics. 
So the model predicts hadronic states with colour charges.

In particular, the $SU(3)_{\rm QCD}$-octet scalar meson, which corresponds to the pseudo-Goldstone
mode, is expected to be rather light compared to the dynamical scale, according to Eq.\,\eqref{octet}.
It is pair produced at the LHC through $SU(3)_{\rm QCD}$ gauge interactions, and it decays into a pair of gluons.
In addition, the octet could be singly produced by gluon fusion (see e.g. \cite{Han:2010rf}), 
via the higher dimensional operator
\begin{eqnarray}\label{SEFT}
{\cal L} \sim \frac{g_s^2}{4\pi \Lambda} S_8 G\tilde{G} \ .
\end{eqnarray}
Here, we have used naive dimensional counting~\cite{Cohen:1997rt,Luty:1997fk}.
From the di-jet searches at the LHC Run-I, the production cross section of an octet scalar with a mass around 
$1$\,TeV is constrained to be ${\cal O}(1)$\,pb~\cite{Aad:2014aqa,Khachatryan:2015sja}.
The Run-I upper limit is much larger than the actual pair production cross section \cite{GoncalvesNetto:2012nt} 
as well as than the single production rate via Eq.~\eqref{SEFT}.%
\footnote{The limit on the octet mass in \cite{Aad:2014aqa} is based on the higher dimensional 
interaction in Eq.\,\eqref{SEFT} with the coefficient $g_s^2/(4\pi\Lambda)$ replaced by $1/m_8$.}
The di-jet searches in Run II have improved the constraint on the production cross section down to ${\cal O}(0.1)$\,~pb,
with which an octet in the TeV range could be testable.  A more detailed, recent analysis can
be found in Ref.~\cite{Belyaev:2016ftv}.

Besides the octet scalars, the model also predicts the stable coloured baryons of Eq.~\eqref{locked_baryons},%
\footnote{Due to the $SU(N_c)_N$ charges of $\psi_{N}$, 
higher dimensional operators which make the coloured baryons decay into the SM particles 
are highly suppressed.
}
Once these stable particles are produced inside colliders,   
they leave visible tracks, whose production cross section
is constrained to be ${\cal O}(10^{-2})$\,pb
for masses in the TeV range~\cite{Khachatryan:2016sfv}.
However, the masses of the baryons are ${\cal O}(N_c\Lambda)$, 
which goes beyond the reach of the LHC for $f \sim \Lambda/4\pi = {\cal O}(1)$\,TeV.
Several other coloured hadrons are predicted, either stable or not, but in any case with masses of ${\cal O}(\Lambda)$ or higher.

In the contact-connection model of section \ref{CCaxion}, with the last module charged only other QCD as in Table \ref{axion2}, 
one predicts elementary, vector-like quarks $Q_{N+1}$, whose masses are of the order of ${\cal O}(\Lambda^3/M_*^2)$.
The vector-like quarks can, for example, decay into the  SM particles via the interactions 
\begin{eqnarray}
{\cal L}_{\rm mix} \sim \epsilon_f   \overline{Q}_{R, N+1} H q_{L,f} + h.c.\ ,
\label{Mixing}
\end{eqnarray}
where $q_L$ and $H$ denote the doublet quark and the Higgs boson in the SM.%
\footnote{These interactions are consistent with the ${\mathbb Z}_{m}^{(j=1\cdots N)}$ symmetry.}
Here, we assume that  the $U(1)_Y$ hypercharge of $\overline{Q}_{R,  N+1}$ is
$-2/3$, corresponding to up-type singlet vector-like quarks. 
We may also assume down-type singlet vector-like quarks by taking the hypercharge to be $1/3$.
The coupling constants, $\epsilon_f$, can be taken to be much smaller than the Yukawa coupling 
constants in the SM and still allow the extra quarks to decay quickly.
For an analysis of electroweak precision constraints on vector-like quarks with generic charges, see Ref.~\cite{Bizot:2015zaa}.
ATLAS and CMS have performed analyses on the production of vector-like quarks 
\cite{ATLAS-CONF-2015-012,ATLAS-CONF-2016-013,CMS-PAS-B2G-12-015,CMS-PAS-B2G-16-002}. 
Assuming that there is no other physics that affects their branching ratios, ATLAS set a $95$\% CL 
lower limit of $m_T > 800$ GeV and $m_B > 735$ GeV from $20.3$\,fb$^{-1}$ of $8$\,TeV 
data, where $T$ is an up-type quark and $B$ is down-type. 
Its analysis for $m_T$ from $3.2$\,fb$^{-1}$ of 13\,TeV  has not surpassed this bound. 
CMS set a $95$\% CL lower limit of $m_T > 750$ from $2.3$\,fb$^{-1}$ of $13$\,TeV data, 
which has already outstripped its result from $8$\,TeV data. 
Looking to the future, CMS expects to rule out an up-type quark of mass less than $1.85$\,TeV at 95\%\,C.L. 
with $3000$\,fb$^{-1}$ of $\sqrt{s}=14$\,TeV data at the HL-LHC \cite{CMS:2013yfa}.

The spectrum of the lightest resonances in the contact-connection model is radically different 
if the last module takes the form of Table \ref{axionComp}.
In this case, the heavy coloured quarks, $Q_N$, are confined within composite states. These have masses of the order of the dynamical scale $\Lambda$ or larger,
with the exception of the pseudo-NGBs. These are a colour octet, whose phenomenology has been already sketched above, as well as colour singlets coupled to the QCD anomaly.
Indeed, in the limit where the contact interaction is small, i.e. $\Lambda \ll M^*$, the $2N$ modes $\pi_j,\xi_j, \, j=1,\dots,N$, defined in section \ref{Model2}, split into $N$ modes $\eta'_j = (\pi_j - q \xi_j)/\sqrt{1+q^2}$,
that receive a mass of order $\Lambda$ from the $SU(N_c)_j$ anomalies, and $N$ massless modes $\pi'_j =  (q \pi_j + \xi_j)/\sqrt{1+q^2}$.
Once the contact interaction is taken into account,  the mass matrix for $\pi'_j$ can be obtained from Eq.~\eqref{EFT2} by integrating out the heavy $\eta'_j$ states.
At leading order in $f/M^*$, one recovers exactly the same mass matrix as the minimal clockwork realisation in Eq.~\eqref{potmin} (without the site $j=0$), 
with the identification $\epsilon \equiv \kappa' (f/M_*)^2\times 2/(1+q^2)$. The matrix diagonalisation \cite{GiudiceMcCullough} 
issues a massless axion $a\equiv a_1$, with couplings to gluons given in Eq.~\eqref{aGG2bis},
and $N-1$ massive pseudo-NGBs, $a_k,\,k=2,\dots,N$, whose masses and couplings to gluons are defined by
\begin{equation}
\label{pionGG}
{\cal L} \simeq \frac 12 \left(\frac{2\kappa' f^4}{M_*^2} \right) \sum_{k=2}^N a_k^2 
+ \frac{g_s^2}{32\pi^2} d({\mathbf R_Q})
\left(\sqrt{\frac{2}{N}} \sum_{k=2}^N (-1)^k \frac{a_k}{f} \sin\frac{(k-1)\pi}{N}\right)
G\tilde{G} 
~,
\end{equation}
where we took the limit $q\gg 1$ for simplicity. These states are close in mass (splitting $\sim 1/q$, see~\cite{GiudiceMcCullough}) and have all couplings to gluons of the same order.
If light enough, they can be produced in gluon fusion and decay back into two gluons, thus the signature is a set of $N-1$ close resonances in di-jet searches.
A recent analysis for one singlet pseudoscalar coupled to gluons can be found in Ref.~\cite{Belyaev:2016ftv}. 
The present LHC bound on the production cross-section is $\mathcal{O}(1)$\,pb for $m_{a_k}=1$\,TeV, which roughly corresponds to $f\gtrsim d({\mathbf R_Q})/5$\,TeV, where we assumed the $N-1$ states are not resolved.

\subsection{Axion dark matter and cosmology}\label{axioncosmo}

The invisible axion is of particular interest as it can be a component of cold dark matter.
The coherent oscillation of the axion  provides a relic axion density \cite{Sikivie:2006ni},
\begin{equation}
\Omega_a h^2 \simeq 0.07 \alpha_i^2 \left(\frac{F_a}{10^{12} \text{ GeV}}\right)^{7/6}\ ,
\end{equation}
where $\alpha_i \in [-\pi, \pi]$ is the initial misalignment angle. 
We assume that PQ symmetry breaking occurs before the primordial inflation,
so that the axion takes a unique field value in the whole observable universe. 
By assuming $\alpha_i ={\cal O}(1)$, the observed dark matter density, 
$\Omega_{DM}h^2 = 0.1197 \pm 0.0022$ \cite{Ade:2015xua}, can be achieved 
for $F_a \sim 10^{12}$\,GeV. 

In the clockwork axion, the effective decay constant is greatly enhanced, $F_a \sim q^N f_a$, 
and hence the required axion dark matter can be achieved by dynamics at the TeV scale.
For example, let us set $f_a = 1$\,TeV.
In the dynamical phase-locking model, $F_a$ is given in Eq.\,\eqref{DynamicalFa} and therefore
the dark matter density is obtained when $N \approx 75$ ($54$)
for $N_c = 4$ ($5$).
These large values are unsurprising, since the value of $q$ is close to one in this model. 
The contact-connection model, on the other hand, produces a larger clockwork factor ($q = -4$ 
in both examples given in section \ref{Model2}). Consequently, the correct relic density is achieved with far fewer sites. 
With the coupling to QCD given in Eq. \eqref{ContactQCDaxion}, for the case 
$N_c = 4, m=2$ ($N_c = 11, m=9$) we find $N \approx 16$ ($N \approx 17$).

Once can search for axion dark matter using a microwave cavity~\cite{Sikivie:1983ip,Bradley:2003kg},
in which the axion is converted to a radio wave with the frequency of the axion mass,
via the axion-photon coupling, 
\begin{eqnarray}
{\cal L} = \frac{C_{a\gamma\gamma}\alpha}{8\pi} \frac{a}{F_a} F_{QED}\tilde F_{QED}\ .
\end{eqnarray}
Here  $C_{a\gamma\gamma}$ is a model dependent coefficient which, for instance, takes the value
\begin{eqnarray}
C_{a\gamma\gamma} =- \frac{2(4+z)}{3(1+z)}\ 
\end{eqnarray}
when one assumes that the extra quarks in the KSVZ models do 
not carry $U(1)_Y$ charges~\cite[for review]{Raffelt:2006cw,Agashe:2014kda}.
Again, $z$ is the ratio of the masses of the up and the down quarks.
So far, the ADMX experiment has put a constraint on the axion-photon coupling of
\begin{eqnarray}
\frac{C_{a\gamma\gamma}\alpha}{2\pi} \frac{1}{F_a}  \lesssim 10^{-15} \,{\rm GeV}^{-1}\ 
\end{eqnarray}
for an axion with a mass of a few $\mu$eV,  assuming that the axion is the dominant component 
of dark matter~\cite{Asztalos:2009yp,Stern:2016bbw}.
The next generation of the ADMX experiment is predicted to search for axion dark matter 
in a mass range up to $40$\,$\mu$eV.

Remarkably, the clockwork axion can be far more visible than in the conventional axion models.
It is possible that the vector-like fermions in the $i$-th clockwork site
carry a $U(1)_Y$ charge of ${\cal O}(1)$.
When $i\ll N$, the axion-photon coupling is enhanced by
\begin{eqnarray}
C_{a\gamma\gamma} \propto q^{N-i} \ ,
\end{eqnarray}
because the contribution to the axion direction from the $i$-th Goldstone mode is $q^{N-i}$ times larger than the one from
the $N$-th mode, which couples to QCD.
Consequently, the clockwork axion can have a much larger coupling to photons than conventional, single-site models.
This of course enhances the detectability of axion dark matter.

Now we turn to the early Universe cosmology of the models. 
As we have discussed in sections \ref{discretesymmetry} and \ref{Model2},
discrete symmetries play an important role in achieving $U(1)_A$ charges in geometric progression. 
Since the discrete symmetries are assumed to be exact, the axion potential generated by the QCD axion 
also respects the discrete symmetries.
The discrete symmetries are spontaneously broken as the temperature 
of the universe goes below the QCD scale.
Thus, if the PQ symmetry breaking takes place after the end of the primordial inflation,
the axion field takes different field values in each Hubble volume at the QCD temperature,
which causes a domain wall problem.
To avoid the domain wall problem, the PQ symmetry breaking is required to take place 
before the primordial inflation, and never to be restored after inflation.%
\footnote{If one allows tiny breaking of the discrete symmetries, the domain walls
can disappear and the domain wall problem is evaded~\cite{Gelmini:1988sf,Larsson:1996sp,Hiramatsu:2012sc}. However, this weakens the theoretical grounds for suppressing the 
quantum gravity effects by exact discrete symmetries.}

We note that the domain wall number can be exponentially large in clockwork axion models
when the clockwork factor, $q$, is fractional.
Let us consider, for example, the contact-connection model of section \ref{Model2}, 
where the phases of the quark-bilinear terms rotate by
\begin{eqnarray}
\arg[\overline{Q}_{Rj} Q_{Lj} ]
\to \arg[\overline{Q}_{Rj} Q_{Lj} ] + q_{Q j} \delta \ , \quad 
\arg[\overline{\psi}_{Rj} \psi_{Lj} ]
\to \arg[\overline{\psi}_{Rj} \psi_{Lj}  ] + q_{\psi j} \delta \ , \quad 
\end{eqnarray}
when the axion field is shifted by
\begin{eqnarray}
\frac{a}{f_a} \to \frac{a}{f_a} +\delta\ .
\end{eqnarray}
Then the domain of the axion is defined by
\begin{eqnarray}
\frac{a}{f_a} \in [0, \delta )\ ,
\end{eqnarray}
where $\delta$ is determined by the requirement that all the factors $q_{Qj,\psi j}\times \delta$ become a multiple of 
$2\pi$ for the first time.
If the clockwork factor is an integer, we find that the domain of the axion is given by
\begin{eqnarray}
\frac{a}{f_a} \in \left[0, q^{N}\times 2\pi\right)\ .
\end{eqnarray}
On the other hand, the axion potential in Eq.\,\eqref{ContactQCDaxion} has a period
\begin{eqnarray}
\frac{a}{f_a} \in [0, q^N/N_f \times 2\pi ) \ .
\end{eqnarray}
Hence, the axion potential possesses
\begin{eqnarray}
\label{DW1}
N_{DW} = N_f\
\end{eqnarray}
degenerate vacua, which determines the domain wall number.
Note that we took $N_f = m$, in order for the model to possess an exact ${\mathbb Z}_m^{N}$ symmetry.
The degeneracy of the vacua reflects an underlying discrete symmetry ${\mathbb Z}_m$, which
is left after the spontaneous breaking of ${\mathbb Z}_m^{N}$.
On the other hand, when the clockwork factor is fractional, the domain wall number becomes exponentially large. 
Indeed, when $q=r/s$, with $r,s$ coprime integers,
the domain of the axion is given by
\begin{eqnarray}
\frac{a}{f_a} \in \left[0, r^{N}\times 2\pi\right)\ .
\end{eqnarray}
Thus, we find that the degeneracy of the vacua is given by
\begin{eqnarray}
N_{DW}  = s^N \times N_f \ ,
\end{eqnarray}
which is exponentially larger than Eq.\,\eqref{DW1}.
This larger degeneracy 
is due to an accidental enhancement of the discrete symmetry of the axion potential,
even if the exact ${\mathbb Z}_m^{N}$ symmetry is spontaneously broken to ${\mathbb Z}_m$ only.
The exponentially large $N_{DW}$ is a generic feature 
of clockwork mechanism models with a fractional clockwork factor.

\subsection{Baryonic dark matter} 

The contact-connection model of section \ref{Model2} also provides the possibility of $SU(N_c)^N$ baryons as dark matter candidates. 
Since coloured stable baryons are strongly constrained cosmologically, 
we focus on the case where strongly-coupled fermions
have no QCD charge, as in Table \ref{axion2}.
Since there are two fermions, $Q_j$ and $\psi_j$, charged under each $SU(N_c)_j$ gauge group for $j=1,\ldots , N$,
there is an exact fermion number symmetry $U(1)_V^{2N}$. Hence,
the $j$-module bound states divide into classes with given charges under $U(1)_{V,Q_j}\times U(1)_{V,\psi_j}$.
Since this symmetry is a generalisation of baryon number, we call baryons all bound states with 
non-vanishing $U(1)_{V,Q_j}\times U(1)_{V,\psi_j}$ charge.
The lightest baryon in each class may be stable, if the decay into two or more baryons of other classes is kinematically forbidden.
In general, one is left with ${\cal O}(2N)$ stable baryons.%
 \footnote{For definiteness, we assume in the following $Q_j^{N_c}$ and $\psi_j^{N_c}$ stable baryons.
In general, stable states may include both  species $Q_j$ and $\psi_j$, and $N_b$ constituent fermions with $N_b\ne N_c$. 
Given the similar annihilation cross-sections,
one expects that the contribution to the relic density is dominated by the heaviest stable baryons, with mass
${\cal O}(N_b\Lambda)$.}

Let us roughly sketch how such baryons can be  dark matter candidates alternative to the axion.
As discussed above, to avoid domain walls we require that the  PQ-symmetry is never restored after inflation,
so the cosmic temperature after inflation is below $\Lambda$.
The baryon masses are $m_{\mathcal{B}} \sim \mathcal{O}(N_c \Lambda)$.
Thus, the thermal relic baryonic dark matter is possible only when the reheating temperature after
inflation is in a small window,%
\footnote{We assume instantaneous reheating at the end of inflation.}
\begin{eqnarray}
\label{temperature}
T_F \lesssim T_R \lesssim  \Lambda\ ,
\end{eqnarray}
where $T_F$ is the freeze-out temperature,
\begin{eqnarray}
T_F \equiv \frac{m_{\mathcal B}}{x_F}\ , \qquad  x_F \sim \frac{1}{2}\log(x_F) + 35 - \log(\frac{m_\mathcal{B}}{\text{TeV}}) \ .
\end{eqnarray}
Here, all the $N$ sites are in thermal equilibrium via scattering with the lighter
pseudo-NGBs $\pi'$'s (see the discussion at the end of subsection \ref{collider}).\footnote{ The heavier $\eta'$ modes decay into three $\pi'$'s or gluons immediately,
while $\pi'$'s decay into QCD jets via Eq.\,\eqref{pionGG} in the model in Table \ref{axionComp}
or into $Q_{N+1}$ in the model in Table \ref{axion2} .
Both can also decay into axions. Thus, each sector does not leave stable particle other than the baryons.}%

Once the reheating temperature is in the window of  Eq.\,\eqref{temperature}, the thermal relic density of the baryonic dark matter 
can be roughly estimated as
\begin{eqnarray}
\Omega_{\cal B} h^2 \sim 0.1 \times \frac{N}{15} \left( \frac{m_{\cal B}}{20\,{\rm TeV}}\right)^2\ ,
\end{eqnarray}
where we assumed that the $2N$ stable baryon states contribute equally and we approximated the annihilation cross section of ${\cal B}$ into $\pi'$'s 
by the unitarity limit~\cite{Griest:1989wd} (see also \cite{Huo:2015nwa}),
\begin{equation}
\left\langle \sigma v \right\rangle \left(\mathcal{B} \overline{\mathcal{B}} \rightarrow \pi' \pi' \right) \sim \frac{4\pi}{m_\mathcal{B}^2}\ .
\label{ann2}
\end{equation}
Thus, in a model with, for example, $N_c \sim10$ and $N \sim 15$, it is possible to explain the observed 
dark matter density by  the thermal relic density of the stable baryons.

\subsection{Further applications}

The clockwork mechanism has also been used in models of natural inflation~\cite{Freese:1990rb} 
and for the relaxion mechanism~\cite{Graham:2015cka}.
The application to the magnetogenesis mechanism in natural inflation models 
can be particularly interesting.
Recently, the existence of cosmological magnetic fields even in void regions
has been suggested by the gamma-ray observations from blazers,
which puts a lower limit on the present magnetic 
field, $B_{\rm eff}\gtrsim 10^{-15}$\,G~\cite{2010Sci...328...73N,2010MNRAS.406L..70T,2011ApJ...727L...4D,Taylor:2011bn,Essey:2010nd,Takahashi:2013lba,Finke:2013tyq,Finke:2015ona}.
Although it is generically difficult to generate such magnetic fields with a long correlation length,
there are arguments that it is possible to achieve it in natural inflation
models where the inflaton is an axion-like particle with 
an anomalous coupling to the $U(1)_Y$ gauge field~\cite{Fujita:2015iga,Adshead:2016iae}.
One needs a hierarchy between the decay constant  appearing in the inflaton potential, 
and the one in the anomalous coupling to the $U(1)_Y$ gauge field.
Such a hierarchy can be easily achieved in the clockwork mechanism,
as we have shown in section \ref{axioncosmo}.


\section{Conclusions}
We have remarked that a sequence of $N$ strongly-coupled sectors may respect some accidental global symmetries
that are collectively carried by the $N$ sites. We focused on the case of an anomaly-free, axial $U(1)_A$ symmetry.
When strong dynamics confines at the dynamical scale $\Lambda$, the $U(1)_A$ symmetry is spontaneously broken
and the decay constant of the associated Goldstone mode, the axion, $a$, is exponentially enhanced at the end of the sequence, 
$F_a \sim q^N f_a$, where $f_a\sim \Lambda/(4\pi)$  and the factor $q>1$ is the ratio between the $U(1)_A$ charges
of neighbouring sectors. 
Clearly, this is an implementation of the clockwork mechanism, which induces a separation between the dynamical scale and the effective interaction
scale of the axion due to a geometric progression of the $U(1)_A$ charges.

In such realisations of the clockwork mechanism, the quantum stability of the dynamical scale 
is guaranteed by dimensional transmutation, as in QCD. As the $N$ gauge groups $SU(N_c)$ have the same matter content,
similar couplings in the ultraviolet nicely imply similar confinement scales. 
As an additional bonus, the occurrence of the global symmetry $U(1)_A$
is enforced by the (discrete) gauge symmetries of the theory, which guarantee that any explicit $U(1)_A$ breaking effect
must be suppressed by a large power of the cutoff.
Indeed, such protection of the axion potential is essential for the most relevant applications, such as the
solution of the strong CP problem by a QCD axion, or inflation via an axion-like scalar field.
We presented three different realisations of the dynamical clockwork, which correspond to specific predictions for $q$,
to distinct sets of new physics states close to the scale $\Lambda$, as well as to different thermal histories.
Let us summarise their main features in turn.

In the {\it phase-locking} model (Table \ref{ModelRepns1}), the link between neighbouring sites is provided by the strong dynamics via fermions charged
under different representations of two adjacent gauge groups. The clockwork factor $q\lesssim 2$ is completely fixed by group theoretical
coefficients, with the upper bound coming from the requirement of asymptotic freedom. 
The sequence configuration is secured by an anomaly-free discrete symmetry ${\mathbb Z}_m^N$, with $m$ also determined by group theory.
When the last site of the clockwork is charged under QCD (Table \ref{ModelRepns1}), 
the $U(1)_A$ Goldstone mode immediately becomes a composite QCD axion with an exponentially large decay constant. The smallness of the $\theta$-parameter is effectively protected from gravity corrections
by the combined effects of the discrete symmetry and of the hierarchy of scales, $\Lambda\ll M_P$.
When $\Lambda$ is close to LHC energies, the model predicts a colour octet of NGBs as well as (heavier) stable coloured hadrons,
and the observed dark matter relic density is obtained from the axion misalignment mechanism for $N\sim 50$. 
On the other hand, since the discrete symmetry implies a large domain wall number, one should not restore $U(1)_A$ 
after the end of inflation, and therefore the reheating temperature should lie below $\Lambda$.

In the {\it contact-connection} model (Table \ref{ModelRepns2}), there are two different representations of fermions
for each strongly-coupled gauge group, and $q$ is the ratio of their respective Dynkin indices, which can be much larger than one, for example 
${\cal O}(N_c)$.
The link between adjacent modules is provided by four-fermion interactions, suppressed by a scale $M_*>\Lambda$.
The configuration of these interactions is secured by charging each pair of adjacent modules under a ${\mathbb Z}_m$ symmetry, free from gauge anomalies.
By adding a last module charged only under QCD (Table \ref{axion2}), one easily realises a composite QCD axion with clockwork-suppressed
couplings. In this case there are no coloured bound states, rather one predicts elementary vector-like quarks
 with mass suppressed by a factor $\Lambda^2/M_*^2$ with respect to $\Lambda$.
The axion relic density may match the dark matter density, for e.g.~$\Lambda \sim 10$\,TeV and $N\sim 10$. 
In addition, there are ${\cal O}(N)$ stable neutral hadrons that also acquire a relic density in the desired range for dark matter, provided that the reheating temperature is just below $\Lambda$ (not larger, in order to avoid domain walls). 
Alternatively, the strong-CP problem can be solved when the last module involves both QCD and the new strong dynamics (Table \ref{axionComp}).
In this case the lightest  states are $N-1$ pion-like mesons with masses suppressed by a factor $\Lambda/M_*$ with respect to $f$, which couple
to gluons through the QCD anomaly.

In the {\it WiFi connection} model (Table \ref{WiFi}), the two fermions of each module transform in the same way under the strong dynamics,
but they have opposite charge under an anomaly-free axial $U(1)$ symmetry.
The link between two adjacent modules is provided by gauging a linear combination of the two $U(1)$ symmetries, 
with $q$ being the arbitrary ratio between the $U(1)$ charges of the two modules. 
In this case, each module contains an exact Goldstone mode, but $N-1$ of those are eaten by the $U(1)$ gauge bosons
and one is left with a massless clockwork axion. Remarkably, the accidental axion $U(1)_A$ is extremely well protected by the sequence of gauged $U(1)$'s:
it can be broken only by operators involving fermions from all the $N$ sectors at the same time.
This phenomenon implies that, even if QCD is coupled to one sector, $U(1)_A$ remains free from the QCD anomaly, and therefore this type of axion cannot 
be used to address the strong $CP$-problem. Still, this WiFi dynamical clockwork emerges as an outstanding mechanism
to generate extremely flat axion potentials.

In summary, we proposed the dynamical clockwork as a flexible mechanism to supply very light scalars with couplings suppressed by a scale
much larger than the strong-coupling scale. We explored several model-building possibilities and applied the mechanism to the QCD axion,
outlining its main phenomenological features. Clearly, several interesting directions are left to investigate,
from the underlying theoretical origin to the connection with models of strongly-coupled electroweak-symmetry breaking, from a more quantitative
study of the QCD axion phenomenology to different cosmological applications of the dynamical, axion-like particles.

\section*{Acknowledgements}
We thank Felix Br\"ummer for useful discussions.
R.C. and M.F. thank Kavli IPMU, Tokyo, where this project was conceived and initiated, for its warm hospitality.
This project has received funding from the European Union's Horizon 2020 research and innovation programme 
under the Marie Sklodowska-Curie grant agreements No 674896 and No 690575 (R.C. and  M.F.).
This work is supported in part by Grants-in-Aid for Scientific Research from the Ministry of Education, Culture, Sports, Science, and Technology (MEXT) KAKENHI, Japan, No.\,25105011, No.\,15H05889 and  No.\,17H02878 (M. I.)
and by the World Premier International Research Center Initiative (WPI), MEXT, Japan (M.I.).


\bibliography{ClockBib_arxiv}

\begin{thebibliography}{97}%
\makeatletter
\providecommand \@ifxundefined [1]{%
 \@ifx{#1\undefined}
}%
\providecommand \@ifnum [1]{%
 \ifnum #1\expandafter \@firstoftwo
 \else \expandafter \@secondoftwo
 \fi
}%
\providecommand \@ifx [1]{%
 \ifx #1\expandafter \@firstoftwo
 \else \expandafter \@secondoftwo
 \fi
}%
\providecommand \natexlab [1]{#1}%
\providecommand \enquote  [1]{``#1''}%
\providecommand \bibnamefont  [1]{#1}%
\providecommand \bibfnamefont [1]{#1}%
\providecommand \citenamefont [1]{#1}%
\providecommand \href@noop [0]{\@secondoftwo}%
\providecommand \href [0]{\begingroup \@sanitize@url \@href}%
\providecommand \@href[1]{\@@startlink{#1}\@@href}%
\providecommand \@@href[1]{\endgroup#1\@@endlink}%
\providecommand \@sanitize@url [0]{\catcode `\\12\catcode `\$12\catcode
  `\&12\catcode `\#12\catcode `\^12\catcode `\_12\catcode `\%12\relax}%
\providecommand \@@startlink[1]{}%
\providecommand \@@endlink[0]{}%
\providecommand \url  [0]{\begingroup\@sanitize@url \@url }%
\providecommand \@url [1]{\endgroup\@href {#1}{\urlprefix }}%
\providecommand \urlprefix  [0]{URL }%
\providecommand \Eprint [0]{\href }%
\providecommand \doibase [0]{http://dx.doi.org/}%
\providecommand \selectlanguage [0]{\@gobble}%
\providecommand \bibinfo  [0]{\@secondoftwo}%
\providecommand \bibfield  [0]{\@secondoftwo}%
\providecommand \translation [1]{[#1]}%
\providecommand \BibitemOpen [0]{}%
\providecommand \bibitemStop [0]{}%
\providecommand \bibitemNoStop [0]{.\EOS\space}%
\providecommand \EOS [0]{\spacefactor3000\relax}%
\providecommand \BibitemShut  [1]{\csname bibitem#1\endcsname}%
\let\auto@bib@innerbib\@empty
\bibitem [{\citenamefont {'t~Hooft}(1980)}]{tHooft:1979rat}%
  \BibitemOpen
  \bibfield  {author} {\bibinfo {author} {\bibfnamefont {G.}~\bibnamefont
  {'t~Hooft}},\ }\bibfield  {booktitle} {\emph {\bibinfo {booktitle} {{Recent
  Developments in Gauge Theories. Proceedings, Nato Advanced Study Institute,
  Cargese, France, August 26 - September 8, 1979}}},\ }\href {\doibase
  10.1007/978-1-4684-7571-5_9} {\bibfield  {journal} {\bibinfo  {journal} {NATO
  Sci. Ser. B}\ }\textbf {\bibinfo {volume} {59}},\ \bibinfo {pages} {135}
  (\bibinfo {year} {1980})}\BibitemShut {NoStop}%
\bibitem [{\citenamefont {Choi}\ and\ \citenamefont {Im}(2016)}]{ChoiIm}%
  \BibitemOpen
  \bibfield  {author} {\bibinfo {author} {\bibfnamefont {K.}~\bibnamefont
  {Choi}}\ and\ \bibinfo {author} {\bibfnamefont {S.~H.}\ \bibnamefont {Im}},\
  }\href {\doibase 10.1007/JHEP01(2016)149} {\bibfield  {journal} {\bibinfo
  {journal} {JHEP}\ }\textbf {\bibinfo {volume} {01}},\ \bibinfo {pages} {149}
  (\bibinfo {year} {2016})},\ \Eprint {http://arxiv.org/abs/1511.00132}
  {arXiv:1511.00132 [hep-ph]} \BibitemShut {NoStop}%
\bibitem [{\citenamefont {Kaplan}\ and\ \citenamefont
  {Rattazzi}(2016)}]{KaplanRattazzi}%
  \BibitemOpen
  \bibfield  {author} {\bibinfo {author} {\bibfnamefont {D.~E.}\ \bibnamefont
  {Kaplan}}\ and\ \bibinfo {author} {\bibfnamefont {R.}~\bibnamefont
  {Rattazzi}},\ }\href {\doibase 10.1103/PhysRevD.93.085007} {\bibfield
  {journal} {\bibinfo  {journal} {Phys. Rev.}\ }\textbf {\bibinfo {volume}
  {D93}},\ \bibinfo {pages} {085007} (\bibinfo {year} {2016})},\ \Eprint
  {http://arxiv.org/abs/1511.01827} {arXiv:1511.01827 [hep-ph]} \BibitemShut
  {NoStop}%
\bibitem [{\citenamefont {Giudice}\ and\ \citenamefont
  {McCullough}(2017{\natexlab{a}})}]{GiudiceMcCullough}%
  \BibitemOpen
  \bibfield  {author} {\bibinfo {author} {\bibfnamefont {G.~F.}\ \bibnamefont
  {Giudice}}\ and\ \bibinfo {author} {\bibfnamefont {M.}~\bibnamefont
  {McCullough}},\ }\href {\doibase 10.1007/JHEP02(2017)036} {\bibfield
  {journal} {\bibinfo  {journal} {JHEP}\ }\textbf {\bibinfo {volume} {02}},\
  \bibinfo {pages} {036} (\bibinfo {year} {2017}{\natexlab{a}})},\ \Eprint
  {http://arxiv.org/abs/1610.07962} {arXiv:1610.07962 [hep-ph]} \BibitemShut
  {NoStop}%
\bibitem [{\citenamefont {Kim}\ \emph {et~al.}(2005)\citenamefont {Kim},
  \citenamefont {Nilles},\ and\ \citenamefont {Peloso}}]{Kim:2004rp}%
  \BibitemOpen
  \bibfield  {author} {\bibinfo {author} {\bibfnamefont {J.~E.}\ \bibnamefont
  {Kim}}, \bibinfo {author} {\bibfnamefont {H.~P.}\ \bibnamefont {Nilles}}, \
  and\ \bibinfo {author} {\bibfnamefont {M.}~\bibnamefont {Peloso}},\ }\href
  {\doibase 10.1088/1475-7516/2005/01/005} {\bibfield  {journal} {\bibinfo
  {journal} {JCAP}\ }\textbf {\bibinfo {volume} {0501}},\ \bibinfo {pages}
  {005} (\bibinfo {year} {2005})},\ \Eprint
  {http://arxiv.org/abs/hep-ph/0409138} {arXiv:hep-ph/0409138 [hep-ph]}
  \BibitemShut {NoStop}%
\bibitem [{\citenamefont {Kehagias}\ and\ \citenamefont
  {Riotto}(2017)}]{KehagiasRiotto}%
  \BibitemOpen
  \bibfield  {author} {\bibinfo {author} {\bibfnamefont {A.}~\bibnamefont
  {Kehagias}}\ and\ \bibinfo {author} {\bibfnamefont {A.}~\bibnamefont
  {Riotto}},\ }\href {\doibase 10.1016/j.physletb.2017.01.042} {\bibfield
  {journal} {\bibinfo  {journal} {Phys. Lett.}\ }\textbf {\bibinfo {volume}
  {B767}},\ \bibinfo {pages} {73} (\bibinfo {year} {2017})},\ \Eprint
  {http://arxiv.org/abs/1611.03316} {arXiv:1611.03316 [hep-ph]} \BibitemShut
  {NoStop}%
\bibitem [{\citenamefont {Farina}\ \emph {et~al.}(2017)\citenamefont {Farina},
  \citenamefont {Pappadopulo}, \citenamefont {Rompineve},\ and\ \citenamefont
  {Tesi}}]{Farina}%
  \BibitemOpen
  \bibfield  {author} {\bibinfo {author} {\bibfnamefont {M.}~\bibnamefont
  {Farina}}, \bibinfo {author} {\bibfnamefont {D.}~\bibnamefont {Pappadopulo}},
  \bibinfo {author} {\bibfnamefont {F.}~\bibnamefont {Rompineve}}, \ and\
  \bibinfo {author} {\bibfnamefont {A.}~\bibnamefont {Tesi}},\ }\href {\doibase
  10.1007/JHEP01(2017)095} {\bibfield  {journal} {\bibinfo  {journal} {JHEP}\
  }\textbf {\bibinfo {volume} {01}},\ \bibinfo {pages} {095} (\bibinfo {year}
  {2017})},\ \Eprint {http://arxiv.org/abs/1611.09855} {arXiv:1611.09855
  [hep-ph]} \BibitemShut {NoStop}%
\bibitem [{\citenamefont {Higaki}\ \emph {et~al.}(2016)\citenamefont {Higaki},
  \citenamefont {Jeong}, \citenamefont {Kitajima},\ and\ \citenamefont
  {Takahashi}}]{Higaki:2016yqk}%
  \BibitemOpen
  \bibfield  {author} {\bibinfo {author} {\bibfnamefont {T.}~\bibnamefont
  {Higaki}}, \bibinfo {author} {\bibfnamefont {K.~S.}\ \bibnamefont {Jeong}},
  \bibinfo {author} {\bibfnamefont {N.}~\bibnamefont {Kitajima}}, \ and\
  \bibinfo {author} {\bibfnamefont {F.}~\bibnamefont {Takahashi}},\ }\href
  {\doibase 10.1007/JHEP06(2016)150} {\bibfield  {journal} {\bibinfo  {journal}
  {JHEP}\ }\textbf {\bibinfo {volume} {06}},\ \bibinfo {pages} {150} (\bibinfo
  {year} {2016})},\ \Eprint {http://arxiv.org/abs/1603.02090} {arXiv:1603.02090
  [hep-ph]} \BibitemShut {NoStop}%
\bibitem [{\citenamefont {Ahmed}\ and\ \citenamefont
  {Dillon}(2016)}]{AhmedDillon}%
  \BibitemOpen
  \bibfield  {author} {\bibinfo {author} {\bibfnamefont {A.}~\bibnamefont
  {Ahmed}}\ and\ \bibinfo {author} {\bibfnamefont {B.~M.}\ \bibnamefont
  {Dillon}},\ }\href@noop {} {\  (\bibinfo {year} {2016})},\ \Eprint
  {http://arxiv.org/abs/1612.04011} {arXiv:1612.04011 [hep-ph]} \BibitemShut
  {NoStop}%
\bibitem [{\citenamefont {Hambye}\ \emph {et~al.}(2016)\citenamefont {Hambye},
  \citenamefont {Teresi},\ and\ \citenamefont {Tytgat}}]{Hambye:2016qkf}%
  \BibitemOpen
  \bibfield  {author} {\bibinfo {author} {\bibfnamefont {T.}~\bibnamefont
  {Hambye}}, \bibinfo {author} {\bibfnamefont {D.}~\bibnamefont {Teresi}}, \
  and\ \bibinfo {author} {\bibfnamefont {M.~H.~G.}\ \bibnamefont {Tytgat}},\
  }\href@noop {} {\  (\bibinfo {year} {2016})},\ \Eprint
  {http://arxiv.org/abs/1612.06411} {arXiv:1612.06411 [hep-ph]} \BibitemShut
  {NoStop}%
\bibitem [{\citenamefont {Graham}\ \emph {et~al.}(2015)\citenamefont {Graham},
  \citenamefont {Kaplan},\ and\ \citenamefont {Rajendran}}]{Graham:2015cka}%
  \BibitemOpen
  \bibfield  {author} {\bibinfo {author} {\bibfnamefont {P.~W.}\ \bibnamefont
  {Graham}}, \bibinfo {author} {\bibfnamefont {D.~E.}\ \bibnamefont {Kaplan}},
  \ and\ \bibinfo {author} {\bibfnamefont {S.}~\bibnamefont {Rajendran}},\
  }\href {\doibase 10.1103/PhysRevLett.115.221801} {\bibfield  {journal}
  {\bibinfo  {journal} {Phys. Rev. Lett.}\ }\textbf {\bibinfo {volume} {115}},\
  \bibinfo {pages} {221801} (\bibinfo {year} {2015})},\ \Eprint
  {http://arxiv.org/abs/1504.07551} {arXiv:1504.07551 [hep-ph]} \BibitemShut
  {NoStop}%
\bibitem [{\citenamefont {Di~Chiara}\ \emph {et~al.}(2016)\citenamefont
  {Di~Chiara}, \citenamefont {Kannike}, \citenamefont {Marzola}, \citenamefont
  {Racioppi}, \citenamefont {Raidal},\ and\ \citenamefont
  {Spethmann}}]{DiChiara:2015euo}%
  \BibitemOpen
  \bibfield  {author} {\bibinfo {author} {\bibfnamefont {S.}~\bibnamefont
  {Di~Chiara}}, \bibinfo {author} {\bibfnamefont {K.}~\bibnamefont {Kannike}},
  \bibinfo {author} {\bibfnamefont {L.}~\bibnamefont {Marzola}}, \bibinfo
  {author} {\bibfnamefont {A.}~\bibnamefont {Racioppi}}, \bibinfo {author}
  {\bibfnamefont {M.}~\bibnamefont {Raidal}}, \ and\ \bibinfo {author}
  {\bibfnamefont {C.}~\bibnamefont {Spethmann}},\ }\href {\doibase
  10.1103/PhysRevD.93.103527} {\bibfield  {journal} {\bibinfo  {journal} {Phys.
  Rev.}\ }\textbf {\bibinfo {volume} {D93}},\ \bibinfo {pages} {103527}
  (\bibinfo {year} {2016})},\ \Eprint {http://arxiv.org/abs/1511.02858}
  {arXiv:1511.02858 [hep-ph]} \BibitemShut {NoStop}%
\bibitem [{\citenamefont {Evans}\ \emph {et~al.}(2016)\citenamefont {Evans},
  \citenamefont {Gherghetta}, \citenamefont {Nagata},\ and\ \citenamefont
  {Thomas}}]{Evans:2016htp}%
  \BibitemOpen
  \bibfield  {author} {\bibinfo {author} {\bibfnamefont {J.~L.}\ \bibnamefont
  {Evans}}, \bibinfo {author} {\bibfnamefont {T.}~\bibnamefont {Gherghetta}},
  \bibinfo {author} {\bibfnamefont {N.}~\bibnamefont {Nagata}}, \ and\ \bibinfo
  {author} {\bibfnamefont {Z.}~\bibnamefont {Thomas}},\ }\href {\doibase
  10.1007/JHEP09(2016)150} {\bibfield  {journal} {\bibinfo  {journal} {JHEP}\
  }\textbf {\bibinfo {volume} {09}},\ \bibinfo {pages} {150} (\bibinfo {year}
  {2016})},\ \Eprint {http://arxiv.org/abs/1602.04812} {arXiv:1602.04812
  [hep-ph]} \BibitemShut {NoStop}%
\bibitem [{\citenamefont {Craig}\ \emph {et~al.}(2017)\citenamefont {Craig},
  \citenamefont {Garcia~Garcia},\ and\ \citenamefont
  {Sutherland}}]{Craig:2017cda}%
  \BibitemOpen
  \bibfield  {author} {\bibinfo {author} {\bibfnamefont {N.}~\bibnamefont
  {Craig}}, \bibinfo {author} {\bibfnamefont {I.}~\bibnamefont
  {Garcia~Garcia}}, \ and\ \bibinfo {author} {\bibfnamefont {D.}~\bibnamefont
  {Sutherland}},\ }\href@noop {} {\  (\bibinfo {year} {2017})},\ \Eprint
  {http://arxiv.org/abs/1704.07831} {arXiv:1704.07831 [hep-ph]} \BibitemShut
  {NoStop}%
\bibitem [{\citenamefont {Giudice}\ and\ \citenamefont
  {McCullough}(2017{\natexlab{b}})}]{Giudice:2017suc}%
  \BibitemOpen
  \bibfield  {author} {\bibinfo {author} {\bibfnamefont {G.~F.}\ \bibnamefont
  {Giudice}}\ and\ \bibinfo {author} {\bibfnamefont {M.}~\bibnamefont
  {McCullough}},\ }\href@noop {} {\  (\bibinfo {year} {2017}{\natexlab{b}})},\
  \Eprint {http://arxiv.org/abs/1705.10162} {arXiv:1705.10162 [hep-ph]}
  \BibitemShut {NoStop}%
\bibitem [{\citenamefont {Harigaya}\ and\ \citenamefont
  {Ibe}(2014{\natexlab{a}})}]{Harigaya:2014eta}%
  \BibitemOpen
  \bibfield  {author} {\bibinfo {author} {\bibfnamefont {K.}~\bibnamefont
  {Harigaya}}\ and\ \bibinfo {author} {\bibfnamefont {M.}~\bibnamefont {Ibe}},\
  }\href {\doibase 10.1016/j.physletb.2014.09.061} {\bibfield  {journal}
  {\bibinfo  {journal} {Phys. Lett.}\ }\textbf {\bibinfo {volume} {B738}},\
  \bibinfo {pages} {301} (\bibinfo {year} {2014}{\natexlab{a}})},\ \Eprint
  {http://arxiv.org/abs/1404.3511} {arXiv:1404.3511 [hep-ph]} \BibitemShut
  {NoStop}%
\bibitem [{\citenamefont {Harigaya}\ and\ \citenamefont
  {Ibe}(2014{\natexlab{b}})}]{Harigaya:2014rga}%
  \BibitemOpen
  \bibfield  {author} {\bibinfo {author} {\bibfnamefont {K.}~\bibnamefont
  {Harigaya}}\ and\ \bibinfo {author} {\bibfnamefont {M.}~\bibnamefont {Ibe}},\
  }\href {\doibase 10.1007/JHEP11(2014)147} {\bibfield  {journal} {\bibinfo
  {journal} {JHEP}\ }\textbf {\bibinfo {volume} {11}},\ \bibinfo {pages} {147}
  (\bibinfo {year} {2014}{\natexlab{b}})},\ \Eprint
  {http://arxiv.org/abs/1407.4893} {arXiv:1407.4893 [hep-ph]} \BibitemShut
  {NoStop}%
\bibitem [{\citenamefont {Raffelt}(2008)}]{Raffelt:2006cw}%
  \BibitemOpen
  \bibfield  {author} {\bibinfo {author} {\bibfnamefont {G.~G.}\ \bibnamefont
  {Raffelt}},\ }\bibfield  {booktitle} {\emph {\bibinfo {booktitle} {{Axions:
  Theory, cosmology, and experimental searches. Proceedings, 1st Joint
  ILIAS-CERN-CAST axion training, Geneva, Switzerland, November 30-December 2,
  2005}}},\ }\href {\doibase 10.1007/978-3-540-73518-2_3} {\bibfield  {journal}
  {\bibinfo  {journal} {Lect. Notes Phys.}\ }\textbf {\bibinfo {volume}
  {741}},\ \bibinfo {pages} {51} (\bibinfo {year} {2008})},\ \bibinfo {note}
  {[,51(2006)]},\ \Eprint {http://arxiv.org/abs/hep-ph/0611350}
  {arXiv:hep-ph/0611350 [hep-ph]} \BibitemShut {NoStop}%
\bibitem [{\citenamefont {Olive}\ \emph {et~al.}(2014)\citenamefont {Olive}
  \emph {et~al.}}]{Agashe:2014kda}%
  \BibitemOpen
  \bibfield  {author} {\bibinfo {author} {\bibfnamefont {K.~A.}\ \bibnamefont
  {Olive}} \emph {et~al.} (\bibinfo {collaboration} {Particle Data Group}),\
  }\href {\doibase 10.1088/1674-1137/38/9/090001} {\bibfield  {journal}
  {\bibinfo  {journal} {Chin. Phys.}\ }\textbf {\bibinfo {volume} {C38}},\
  \bibinfo {pages} {090001} (\bibinfo {year} {2014})}\BibitemShut {NoStop}%
\bibitem [{\citenamefont {Kim}(1979)}]{Kim:1979if}%
  \BibitemOpen
  \bibfield  {author} {\bibinfo {author} {\bibfnamefont {J.~E.}\ \bibnamefont
  {Kim}},\ }\href {\doibase 10.1103/PhysRevLett.43.103} {\bibfield  {journal}
  {\bibinfo  {journal} {Phys. Rev. Lett.}\ }\textbf {\bibinfo {volume} {43}},\
  \bibinfo {pages} {103} (\bibinfo {year} {1979})}\BibitemShut {NoStop}%
\bibitem [{\citenamefont {Shifman}\ \emph {et~al.}(1980)\citenamefont
  {Shifman}, \citenamefont {Vainshtein},\ and\ \citenamefont
  {Zakharov}}]{Shifman:1979if}%
  \BibitemOpen
  \bibfield  {author} {\bibinfo {author} {\bibfnamefont {M.~A.}\ \bibnamefont
  {Shifman}}, \bibinfo {author} {\bibfnamefont {A.~I.}\ \bibnamefont
  {Vainshtein}}, \ and\ \bibinfo {author} {\bibfnamefont {V.~I.}\ \bibnamefont
  {Zakharov}},\ }\href {\doibase 10.1016/0550-3213(80)90209-6} {\bibfield
  {journal} {\bibinfo  {journal} {Nucl. Phys.}\ }\textbf {\bibinfo {volume}
  {B166}},\ \bibinfo {pages} {493} (\bibinfo {year} {1980})}\BibitemShut
  {NoStop}%
\bibitem [{\citenamefont {Kim}(1985)}]{Kim:1984pt}%
  \BibitemOpen
  \bibfield  {author} {\bibinfo {author} {\bibfnamefont {J.~E.}\ \bibnamefont
  {Kim}},\ }\href {\doibase 10.1103/PhysRevD.31.1733} {\bibfield  {journal}
  {\bibinfo  {journal} {Phys. Rev.}\ }\textbf {\bibinfo {volume} {D31}},\
  \bibinfo {pages} {1733} (\bibinfo {year} {1985})}\BibitemShut {NoStop}%
\bibitem [{\citenamefont {Barnard}\ \emph {et~al.}(2014)\citenamefont
  {Barnard}, \citenamefont {Gherghetta},\ and\ \citenamefont
  {Ray}}]{Barnard:2013zea}%
  \BibitemOpen
  \bibfield  {author} {\bibinfo {author} {\bibfnamefont {J.}~\bibnamefont
  {Barnard}}, \bibinfo {author} {\bibfnamefont {T.}~\bibnamefont {Gherghetta}},
  \ and\ \bibinfo {author} {\bibfnamefont {T.~S.}\ \bibnamefont {Ray}},\ }\href
  {\doibase 10.1007/JHEP02(2014)002} {\bibfield  {journal} {\bibinfo  {journal}
  {JHEP}\ }\textbf {\bibinfo {volume} {02}},\ \bibinfo {pages} {002} (\bibinfo
  {year} {2014})},\ \Eprint {http://arxiv.org/abs/1311.6562} {arXiv:1311.6562
  [hep-ph]} \BibitemShut {NoStop}%
\bibitem [{\citenamefont {Ferretti}\ and\ \citenamefont
  {Karateev}(2014)}]{Ferretti:2013kya}%
  \BibitemOpen
  \bibfield  {author} {\bibinfo {author} {\bibfnamefont {G.}~\bibnamefont
  {Ferretti}}\ and\ \bibinfo {author} {\bibfnamefont {D.}~\bibnamefont
  {Karateev}},\ }\href {\doibase 10.1007/JHEP03(2014)077} {\bibfield  {journal}
  {\bibinfo  {journal} {JHEP}\ }\textbf {\bibinfo {volume} {03}},\ \bibinfo
  {pages} {077} (\bibinfo {year} {2014})},\ \Eprint
  {http://arxiv.org/abs/1312.5330} {arXiv:1312.5330 [hep-ph]} \BibitemShut
  {NoStop}%
\bibitem [{\citenamefont {Cacciapaglia}\ and\ \citenamefont
  {Sannino}(2014)}]{Cacciapaglia:2014uja}%
  \BibitemOpen
  \bibfield  {author} {\bibinfo {author} {\bibfnamefont {G.}~\bibnamefont
  {Cacciapaglia}}\ and\ \bibinfo {author} {\bibfnamefont {F.}~\bibnamefont
  {Sannino}},\ }\href {\doibase 10.1007/JHEP04(2014)111} {\bibfield  {journal}
  {\bibinfo  {journal} {JHEP}\ }\textbf {\bibinfo {volume} {04}},\ \bibinfo
  {pages} {111} (\bibinfo {year} {2014})},\ \Eprint
  {http://arxiv.org/abs/1402.0233} {arXiv:1402.0233 [hep-ph]} \BibitemShut
  {NoStop}%
\bibitem [{\citenamefont {Vecchi}(2017)}]{Vecchi:2015fma}%
  \BibitemOpen
  \bibfield  {author} {\bibinfo {author} {\bibfnamefont {L.}~\bibnamefont
  {Vecchi}},\ }\href {\doibase 10.1007/JHEP02(2017)094} {\bibfield  {journal}
  {\bibinfo  {journal} {JHEP}\ }\textbf {\bibinfo {volume} {02}},\ \bibinfo
  {pages} {094} (\bibinfo {year} {2017})},\ \Eprint
  {http://arxiv.org/abs/1506.00623} {arXiv:1506.00623 [hep-ph]} \BibitemShut
  {NoStop}%
\bibitem [{\citenamefont {Bizot}\ \emph {et~al.}(2017)\citenamefont {Bizot},
  \citenamefont {Frigerio}, \citenamefont {Knecht},\ and\ \citenamefont
  {Kneur}}]{Bizot:2016zyu}%
  \BibitemOpen
  \bibfield  {author} {\bibinfo {author} {\bibfnamefont {N.}~\bibnamefont
  {Bizot}}, \bibinfo {author} {\bibfnamefont {M.}~\bibnamefont {Frigerio}},
  \bibinfo {author} {\bibfnamefont {M.}~\bibnamefont {Knecht}}, \ and\ \bibinfo
  {author} {\bibfnamefont {J.-L.}\ \bibnamefont {Kneur}},\ }\href {\doibase
  10.1103/PhysRevD.95.075006} {\bibfield  {journal} {\bibinfo  {journal} {Phys.
  Rev.}\ }\textbf {\bibinfo {volume} {D95}},\ \bibinfo {pages} {075006}
  (\bibinfo {year} {2017})},\ \Eprint {http://arxiv.org/abs/1610.09293}
  {arXiv:1610.09293 [hep-ph]} \BibitemShut {NoStop}%
\bibitem [{\citenamefont {Hawking}(1987)}]{Hawking:1987mz}%
  \BibitemOpen
  \bibfield  {author} {\bibinfo {author} {\bibfnamefont {S.~W.}\ \bibnamefont
  {Hawking}},\ }\bibfield  {booktitle} {\emph {\bibinfo {booktitle} {{Moscow
  Quantum Grav.1987:0125}}},\ }\href {\doibase 10.1016/0370-2693(87)90028-1}
  {\bibfield  {journal} {\bibinfo  {journal} {Phys. Lett.}\ }\textbf {\bibinfo
  {volume} {B195}},\ \bibinfo {pages} {337} (\bibinfo {year}
  {1987})}\BibitemShut {NoStop}%
\bibitem [{\citenamefont {Lavrelashvili}\ \emph {et~al.}(1987)\citenamefont
  {Lavrelashvili}, \citenamefont {Rubakov},\ and\ \citenamefont
  {Tinyakov}}]{Lavrelashvili:1987jg}%
  \BibitemOpen
  \bibfield  {author} {\bibinfo {author} {\bibfnamefont {G.~V.}\ \bibnamefont
  {Lavrelashvili}}, \bibinfo {author} {\bibfnamefont {V.~A.}\ \bibnamefont
  {Rubakov}}, \ and\ \bibinfo {author} {\bibfnamefont {P.~G.}\ \bibnamefont
  {Tinyakov}},\ }\href@noop {} {\bibfield  {journal} {\bibinfo  {journal} {JETP
  Lett.}\ }\textbf {\bibinfo {volume} {46}},\ \bibinfo {pages} {167} (\bibinfo
  {year} {1987})},\ \bibinfo {note} {[Pisma Zh. Eksp. Teor.
  Fiz.46,134(1987)]}\BibitemShut {NoStop}%
\bibitem [{\citenamefont {Giddings}\ and\ \citenamefont
  {Strominger}(1988)}]{Giddings:1988cx}%
  \BibitemOpen
  \bibfield  {author} {\bibinfo {author} {\bibfnamefont {S.~B.}\ \bibnamefont
  {Giddings}}\ and\ \bibinfo {author} {\bibfnamefont {A.}~\bibnamefont
  {Strominger}},\ }\href {\doibase 10.1016/0550-3213(88)90109-5} {\bibfield
  {journal} {\bibinfo  {journal} {Nucl. Phys.}\ }\textbf {\bibinfo {volume}
  {B307}},\ \bibinfo {pages} {854} (\bibinfo {year} {1988})}\BibitemShut
  {NoStop}%
\bibitem [{\citenamefont {Coleman}(1988)}]{Coleman:1988tj}%
  \BibitemOpen
  \bibfield  {author} {\bibinfo {author} {\bibfnamefont {S.~R.}\ \bibnamefont
  {Coleman}},\ }\href {\doibase 10.1016/0550-3213(88)90097-1} {\bibfield
  {journal} {\bibinfo  {journal} {Nucl. Phys.}\ }\textbf {\bibinfo {volume}
  {B310}},\ \bibinfo {pages} {643} (\bibinfo {year} {1988})}\BibitemShut
  {NoStop}%
\bibitem [{\citenamefont {Gilbert}(1989)}]{Gilbert:1989nq}%
  \BibitemOpen
  \bibfield  {author} {\bibinfo {author} {\bibfnamefont {G.}~\bibnamefont
  {Gilbert}},\ }\href {\doibase 10.1016/0550-3213(89)90097-7} {\bibfield
  {journal} {\bibinfo  {journal} {Nucl. Phys.}\ }\textbf {\bibinfo {volume}
  {B328}},\ \bibinfo {pages} {159} (\bibinfo {year} {1989})}\BibitemShut
  {NoStop}%
\bibitem [{\citenamefont {Kim}(1981)}]{Kim:1981bb}%
  \BibitemOpen
  \bibfield  {author} {\bibinfo {author} {\bibfnamefont {J.~E.}\ \bibnamefont
  {Kim}},\ }\href {\doibase 10.1103/PhysRevD.24.3007} {\bibfield  {journal}
  {\bibinfo  {journal} {Phys. Rev.}\ }\textbf {\bibinfo {volume} {D24}},\
  \bibinfo {pages} {3007} (\bibinfo {year} {1981})}\BibitemShut {NoStop}%
\bibitem [{\citenamefont {Georgi}\ \emph {et~al.}(1981)\citenamefont {Georgi},
  \citenamefont {Hall},\ and\ \citenamefont {Wise}}]{Georgi:1981pu}%
  \BibitemOpen
  \bibfield  {author} {\bibinfo {author} {\bibfnamefont {H.~M.}\ \bibnamefont
  {Georgi}}, \bibinfo {author} {\bibfnamefont {L.~J.}\ \bibnamefont {Hall}}, \
  and\ \bibinfo {author} {\bibfnamefont {M.~B.}\ \bibnamefont {Wise}},\ }\href
  {\doibase 10.1016/0550-3213(81)90433-8} {\bibfield  {journal} {\bibinfo
  {journal} {Nucl. Phys.}\ }\textbf {\bibinfo {volume} {B192}},\ \bibinfo
  {pages} {409} (\bibinfo {year} {1981})}\BibitemShut {NoStop}%
\bibitem [{\citenamefont {Dimopoulos}\ \emph {et~al.}(1982)\citenamefont
  {Dimopoulos}, \citenamefont {Frampton}, \citenamefont {Georgi},\ and\
  \citenamefont {Wise}}]{Dimopoulos:1982my}%
  \BibitemOpen
  \bibfield  {author} {\bibinfo {author} {\bibfnamefont {S.}~\bibnamefont
  {Dimopoulos}}, \bibinfo {author} {\bibfnamefont {P.~H.}\ \bibnamefont
  {Frampton}}, \bibinfo {author} {\bibfnamefont {H.}~\bibnamefont {Georgi}}, \
  and\ \bibinfo {author} {\bibfnamefont {M.~B.}\ \bibnamefont {Wise}},\ }\href
  {\doibase 10.1016/0370-2693(82)90543-3} {\bibfield  {journal} {\bibinfo
  {journal} {Phys. Lett.}\ }\textbf {\bibinfo {volume} {B117}},\ \bibinfo
  {pages} {185} (\bibinfo {year} {1982})}\BibitemShut {NoStop}%
\bibitem [{\citenamefont {Frampton}(1982)}]{Frampton:1981qu}%
  \BibitemOpen
  \bibfield  {author} {\bibinfo {author} {\bibfnamefont {P.~H.}\ \bibnamefont
  {Frampton}},\ }\href {\doibase 10.1103/PhysRevD.25.294} {\bibfield  {journal}
  {\bibinfo  {journal} {Phys. Rev.}\ }\textbf {\bibinfo {volume} {D25}},\
  \bibinfo {pages} {294} (\bibinfo {year} {1982})}\BibitemShut {NoStop}%
\bibitem [{\citenamefont {Kang}\ \emph {et~al.}(1982)\citenamefont {Kang},
  \citenamefont {Koh},\ and\ \citenamefont {Ouvry}}]{Kang:1982bx}%
  \BibitemOpen
  \bibfield  {author} {\bibinfo {author} {\bibfnamefont {K.}~\bibnamefont
  {Kang}}, \bibinfo {author} {\bibfnamefont {I.-G.}\ \bibnamefont {Koh}}, \
  and\ \bibinfo {author} {\bibfnamefont {S.}~\bibnamefont {Ouvry}},\ }\href
  {\doibase 10.1016/0370-2693(82)90689-X} {\bibfield  {journal} {\bibinfo
  {journal} {Phys. Lett.}\ }\textbf {\bibinfo {volume} {B119}},\ \bibinfo
  {pages} {361} (\bibinfo {year} {1982})}\BibitemShut {NoStop}%
\bibitem [{\citenamefont {Lazarides}\ and\ \citenamefont
  {Shafi}(1982)}]{Lazarides:1982tw}%
  \BibitemOpen
  \bibfield  {author} {\bibinfo {author} {\bibfnamefont {G.}~\bibnamefont
  {Lazarides}}\ and\ \bibinfo {author} {\bibfnamefont {Q.}~\bibnamefont
  {Shafi}},\ }\href {\doibase 10.1016/0370-2693(82)90506-8} {\bibfield
  {journal} {\bibinfo  {journal} {Phys. Lett.}\ }\textbf {\bibinfo {volume}
  {B115}},\ \bibinfo {pages} {21} (\bibinfo {year} {1982})}\BibitemShut
  {NoStop}%
\bibitem [{\citenamefont {Banks}\ \emph {et~al.}(1991)\citenamefont {Banks},
  \citenamefont {Dine},\ and\ \citenamefont {Seiberg}}]{Banks:1991mb}%
  \BibitemOpen
  \bibfield  {author} {\bibinfo {author} {\bibfnamefont {T.}~\bibnamefont
  {Banks}}, \bibinfo {author} {\bibfnamefont {M.}~\bibnamefont {Dine}}, \ and\
  \bibinfo {author} {\bibfnamefont {N.}~\bibnamefont {Seiberg}},\ }\href
  {\doibase 10.1016/0370-2693(91)90561-4} {\bibfield  {journal} {\bibinfo
  {journal} {Phys. Lett.}\ }\textbf {\bibinfo {volume} {B273}},\ \bibinfo
  {pages} {105} (\bibinfo {year} {1991})},\ \Eprint
  {http://arxiv.org/abs/hep-th/9109040} {arXiv:hep-th/9109040 [hep-th]}
  \BibitemShut {NoStop}%
\bibitem [{\citenamefont {Barr}\ and\ \citenamefont
  {Seckel}(1992)}]{Barr:1992qq}%
  \BibitemOpen
  \bibfield  {author} {\bibinfo {author} {\bibfnamefont {S.~M.}\ \bibnamefont
  {Barr}}\ and\ \bibinfo {author} {\bibfnamefont {D.}~\bibnamefont {Seckel}},\
  }\href {\doibase 10.1103/PhysRevD.46.539} {\bibfield  {journal} {\bibinfo
  {journal} {Phys. Rev.}\ }\textbf {\bibinfo {volume} {D46}},\ \bibinfo {pages}
  {539} (\bibinfo {year} {1992})}\BibitemShut {NoStop}%
\bibitem [{\citenamefont {Kamionkowski}\ and\ \citenamefont
  {March-Russell}(1992)}]{Kamionkowski:1992mf}%
  \BibitemOpen
  \bibfield  {author} {\bibinfo {author} {\bibfnamefont {M.}~\bibnamefont
  {Kamionkowski}}\ and\ \bibinfo {author} {\bibfnamefont {J.}~\bibnamefont
  {March-Russell}},\ }\href {\doibase 10.1016/0370-2693(92)90492-M} {\bibfield
  {journal} {\bibinfo  {journal} {Phys. Lett.}\ }\textbf {\bibinfo {volume}
  {B282}},\ \bibinfo {pages} {137} (\bibinfo {year} {1992})},\ \Eprint
  {http://arxiv.org/abs/hep-th/9202003} {arXiv:hep-th/9202003 [hep-th]}
  \BibitemShut {NoStop}%
\bibitem [{\citenamefont {Holman}\ \emph {et~al.}(1992)\citenamefont {Holman},
  \citenamefont {Hsu}, \citenamefont {Kephart}, \citenamefont {Kolb},
  \citenamefont {Watkins},\ and\ \citenamefont {Widrow}}]{Holman:1992us}%
  \BibitemOpen
  \bibfield  {author} {\bibinfo {author} {\bibfnamefont {R.}~\bibnamefont
  {Holman}}, \bibinfo {author} {\bibfnamefont {S.~D.~H.}\ \bibnamefont {Hsu}},
  \bibinfo {author} {\bibfnamefont {T.~W.}\ \bibnamefont {Kephart}}, \bibinfo
  {author} {\bibfnamefont {E.~W.}\ \bibnamefont {Kolb}}, \bibinfo {author}
  {\bibfnamefont {R.}~\bibnamefont {Watkins}}, \ and\ \bibinfo {author}
  {\bibfnamefont {L.~M.}\ \bibnamefont {Widrow}},\ }\href {\doibase
  10.1016/0370-2693(92)90491-L} {\bibfield  {journal} {\bibinfo  {journal}
  {Phys. Lett.}\ }\textbf {\bibinfo {volume} {B282}},\ \bibinfo {pages} {132}
  (\bibinfo {year} {1992})},\ \Eprint {http://arxiv.org/abs/hep-ph/9203206}
  {arXiv:hep-ph/9203206 [hep-ph]} \BibitemShut {NoStop}%
\bibitem [{\citenamefont {Dine}(1992)}]{Dine:1992vx}%
  \BibitemOpen
  \bibfield  {author} {\bibinfo {author} {\bibfnamefont {M.}~\bibnamefont
  {Dine}},\ }in\ \href
  {http://inspirehep.net/record/32647/files/arXiv:hep-th_9207045.pdf} {\emph
  {\bibinfo {booktitle} {{Conference on Topics in Quantum Gravity Cincinnati,
  Ohio, April 3-4, 1992}}}}\ (\bibinfo {year} {1992})\ pp.\ \bibinfo {pages}
  {157--169},\ \Eprint {http://arxiv.org/abs/hep-th/9207045}
  {arXiv:hep-th/9207045 [hep-th]} \BibitemShut {NoStop}%
\bibitem [{\citenamefont {Dias}\ \emph {et~al.}(2003)\citenamefont {Dias},
  \citenamefont {Pleitez},\ and\ \citenamefont {Tonasse}}]{Dias:2002gg}%
  \BibitemOpen
  \bibfield  {author} {\bibinfo {author} {\bibfnamefont {A.~G.}\ \bibnamefont
  {Dias}}, \bibinfo {author} {\bibfnamefont {V.}~\bibnamefont {Pleitez}}, \
  and\ \bibinfo {author} {\bibfnamefont {M.~D.}\ \bibnamefont {Tonasse}},\
  }\href {\doibase 10.1103/PhysRevD.67.095008} {\bibfield  {journal} {\bibinfo
  {journal} {Phys. Rev.}\ }\textbf {\bibinfo {volume} {D67}},\ \bibinfo {pages}
  {095008} (\bibinfo {year} {2003})},\ \Eprint
  {http://arxiv.org/abs/hep-ph/0211107} {arXiv:hep-ph/0211107 [hep-ph]}
  \BibitemShut {NoStop}%
\bibitem [{\citenamefont {Dvali}(2005)}]{Dvali:2005an}%
  \BibitemOpen
  \bibfield  {author} {\bibinfo {author} {\bibfnamefont {G.}~\bibnamefont
  {Dvali}},\ }\href@noop {} {\  (\bibinfo {year} {2005})},\ \Eprint
  {http://arxiv.org/abs/hep-th/0507215} {arXiv:hep-th/0507215 [hep-th]}
  \BibitemShut {NoStop}%
\bibitem [{\citenamefont {Carpenter}\ \emph {et~al.}(2009)\citenamefont
  {Carpenter}, \citenamefont {Dine},\ and\ \citenamefont
  {Festuccia}}]{Carpenter:2009zs}%
  \BibitemOpen
  \bibfield  {author} {\bibinfo {author} {\bibfnamefont {L.~M.}\ \bibnamefont
  {Carpenter}}, \bibinfo {author} {\bibfnamefont {M.}~\bibnamefont {Dine}}, \
  and\ \bibinfo {author} {\bibfnamefont {G.}~\bibnamefont {Festuccia}},\ }\href
  {\doibase 10.1103/PhysRevD.80.125017} {\bibfield  {journal} {\bibinfo
  {journal} {Phys. Rev.}\ }\textbf {\bibinfo {volume} {D80}},\ \bibinfo {pages}
  {125017} (\bibinfo {year} {2009})},\ \Eprint {http://arxiv.org/abs/0906.1273}
  {arXiv:0906.1273 [hep-th]} \BibitemShut {NoStop}%
\bibitem [{\citenamefont {Harigaya}\ \emph {et~al.}(2013)\citenamefont
  {Harigaya}, \citenamefont {Ibe}, \citenamefont {Schmitz},\ and\ \citenamefont
  {Yanagida}}]{Harigaya:2013vja}%
  \BibitemOpen
  \bibfield  {author} {\bibinfo {author} {\bibfnamefont {K.}~\bibnamefont
  {Harigaya}}, \bibinfo {author} {\bibfnamefont {M.}~\bibnamefont {Ibe}},
  \bibinfo {author} {\bibfnamefont {K.}~\bibnamefont {Schmitz}}, \ and\
  \bibinfo {author} {\bibfnamefont {T.~T.}\ \bibnamefont {Yanagida}},\ }\href
  {\doibase 10.1103/PhysRevD.88.075022} {\bibfield  {journal} {\bibinfo
  {journal} {Phys. Rev.}\ }\textbf {\bibinfo {volume} {D88}},\ \bibinfo {pages}
  {075022} (\bibinfo {year} {2013})},\ \Eprint {http://arxiv.org/abs/1308.1227}
  {arXiv:1308.1227 [hep-ph]} \BibitemShut {NoStop}%
\bibitem [{\citenamefont {Harigaya}\ \emph {et~al.}(2015)\citenamefont
  {Harigaya}, \citenamefont {Ibe}, \citenamefont {Schmitz},\ and\ \citenamefont
  {Yanagida}}]{Harigaya:2015soa}%
  \BibitemOpen
  \bibfield  {author} {\bibinfo {author} {\bibfnamefont {K.}~\bibnamefont
  {Harigaya}}, \bibinfo {author} {\bibfnamefont {M.}~\bibnamefont {Ibe}},
  \bibinfo {author} {\bibfnamefont {K.}~\bibnamefont {Schmitz}}, \ and\
  \bibinfo {author} {\bibfnamefont {T.~T.}\ \bibnamefont {Yanagida}},\ }\href
  {\doibase 10.1103/PhysRevD.92.075003} {\bibfield  {journal} {\bibinfo
  {journal} {Phys. Rev.}\ }\textbf {\bibinfo {volume} {D92}},\ \bibinfo {pages}
  {075003} (\bibinfo {year} {2015})},\ \Eprint
  {http://arxiv.org/abs/1505.07388} {arXiv:1505.07388 [hep-ph]} \BibitemShut
  {NoStop}%
\bibitem [{\citenamefont {Redi}\ and\ \citenamefont
  {Sato}(2016)}]{Redi:2016esr}%
  \BibitemOpen
  \bibfield  {author} {\bibinfo {author} {\bibfnamefont {M.}~\bibnamefont
  {Redi}}\ and\ \bibinfo {author} {\bibfnamefont {R.}~\bibnamefont {Sato}},\
  }\href {\doibase 10.1007/JHEP05(2016)104} {\bibfield  {journal} {\bibinfo
  {journal} {JHEP}\ }\textbf {\bibinfo {volume} {05}},\ \bibinfo {pages} {104}
  (\bibinfo {year} {2016})},\ \Eprint {http://arxiv.org/abs/1602.05427}
  {arXiv:1602.05427 [hep-ph]} \BibitemShut {NoStop}%
\bibitem [{\citenamefont {Fukuda}\ \emph {et~al.}(2017)\citenamefont {Fukuda},
  \citenamefont {Ibe}, \citenamefont {Suzuki},\ and\ \citenamefont
  {Yanagida}}]{Fukuda:2017ylt}%
  \BibitemOpen
  \bibfield  {author} {\bibinfo {author} {\bibfnamefont {H.}~\bibnamefont
  {Fukuda}}, \bibinfo {author} {\bibfnamefont {M.}~\bibnamefont {Ibe}},
  \bibinfo {author} {\bibfnamefont {M.}~\bibnamefont {Suzuki}}, \ and\ \bibinfo
  {author} {\bibfnamefont {T.~T.}\ \bibnamefont {Yanagida}},\ }\href@noop {} {\
   (\bibinfo {year} {2017})},\ \Eprint {http://arxiv.org/abs/1703.01112}
  {arXiv:1703.01112 [hep-ph]} \BibitemShut {NoStop}%
\bibitem [{\citenamefont {Banks}\ and\ \citenamefont
  {Seiberg}(2011)}]{Banks:2010zn}%
  \BibitemOpen
  \bibfield  {author} {\bibinfo {author} {\bibfnamefont {T.}~\bibnamefont
  {Banks}}\ and\ \bibinfo {author} {\bibfnamefont {N.}~\bibnamefont
  {Seiberg}},\ }\href {\doibase 10.1103/PhysRevD.83.084019} {\bibfield
  {journal} {\bibinfo  {journal} {Phys. Rev.}\ }\textbf {\bibinfo {volume}
  {D83}},\ \bibinfo {pages} {084019} (\bibinfo {year} {2011})},\ \Eprint
  {http://arxiv.org/abs/1011.5120} {arXiv:1011.5120 [hep-th]} \BibitemShut
  {NoStop}%
\bibitem [{\citenamefont {Krauss}\ and\ \citenamefont
  {Wilczek}(1989)}]{Krauss:1988zc}%
  \BibitemOpen
  \bibfield  {author} {\bibinfo {author} {\bibfnamefont {L.~M.}\ \bibnamefont
  {Krauss}}\ and\ \bibinfo {author} {\bibfnamefont {F.}~\bibnamefont
  {Wilczek}},\ }\href {\doibase 10.1103/PhysRevLett.62.1221} {\bibfield
  {journal} {\bibinfo  {journal} {Phys. Rev. Lett.}\ }\textbf {\bibinfo
  {volume} {62}},\ \bibinfo {pages} {1221} (\bibinfo {year}
  {1989})}\BibitemShut {NoStop}%
\bibitem [{\citenamefont {Preskill}\ and\ \citenamefont
  {Krauss}(1990)}]{Preskill:1990bm}%
  \BibitemOpen
  \bibfield  {author} {\bibinfo {author} {\bibfnamefont {J.}~\bibnamefont
  {Preskill}}\ and\ \bibinfo {author} {\bibfnamefont {L.~M.}\ \bibnamefont
  {Krauss}},\ }\href {\doibase 10.1016/0550-3213(90)90262-C} {\bibfield
  {journal} {\bibinfo  {journal} {Nucl. Phys.}\ }\textbf {\bibinfo {volume}
  {B341}},\ \bibinfo {pages} {50} (\bibinfo {year} {1990})}\BibitemShut
  {NoStop}%
\bibitem [{\citenamefont {Preskill}\ \emph {et~al.}(1991)\citenamefont
  {Preskill}, \citenamefont {Trivedi}, \citenamefont {Wilczek},\ and\
  \citenamefont {Wise}}]{Preskill:1991kd}%
  \BibitemOpen
  \bibfield  {author} {\bibinfo {author} {\bibfnamefont {J.}~\bibnamefont
  {Preskill}}, \bibinfo {author} {\bibfnamefont {S.~P.}\ \bibnamefont
  {Trivedi}}, \bibinfo {author} {\bibfnamefont {F.}~\bibnamefont {Wilczek}}, \
  and\ \bibinfo {author} {\bibfnamefont {M.~B.}\ \bibnamefont {Wise}},\ }\href
  {\doibase 10.1016/0550-3213(91)90241-O} {\bibfield  {journal} {\bibinfo
  {journal} {Nucl. Phys.}\ }\textbf {\bibinfo {volume} {B363}},\ \bibinfo
  {pages} {207} (\bibinfo {year} {1991})}\BibitemShut {NoStop}%
\bibitem [{\citenamefont {Banks}\ and\ \citenamefont
  {Dine}(1992)}]{Banks:1991xj}%
  \BibitemOpen
  \bibfield  {author} {\bibinfo {author} {\bibfnamefont {T.}~\bibnamefont
  {Banks}}\ and\ \bibinfo {author} {\bibfnamefont {M.}~\bibnamefont {Dine}},\
  }\href {\doibase 10.1103/PhysRevD.45.1424} {\bibfield  {journal} {\bibinfo
  {journal} {Phys. Rev.}\ }\textbf {\bibinfo {volume} {D45}},\ \bibinfo {pages}
  {1424} (\bibinfo {year} {1992})},\ \Eprint
  {http://arxiv.org/abs/hep-th/9109045} {arXiv:hep-th/9109045 [hep-th]}
  \BibitemShut {NoStop}%
\bibitem [{\citenamefont {Peccei}\ and\ \citenamefont
  {Quinn}(1977{\natexlab{a}})}]{Peccei:1977hh}%
  \BibitemOpen
  \bibfield  {author} {\bibinfo {author} {\bibfnamefont {R.~D.}\ \bibnamefont
  {Peccei}}\ and\ \bibinfo {author} {\bibfnamefont {H.~R.}\ \bibnamefont
  {Quinn}},\ }\href {\doibase 10.1103/PhysRevLett.38.1440} {\bibfield
  {journal} {\bibinfo  {journal} {Phys. Rev. Lett.}\ }\textbf {\bibinfo
  {volume} {38}},\ \bibinfo {pages} {1440} (\bibinfo {year}
  {1977}{\natexlab{a}})}\BibitemShut {NoStop}%
\bibitem [{\citenamefont {Peccei}\ and\ \citenamefont
  {Quinn}(1977{\natexlab{b}})}]{Peccei:1977ur}%
  \BibitemOpen
  \bibfield  {author} {\bibinfo {author} {\bibfnamefont {R.~D.}\ \bibnamefont
  {Peccei}}\ and\ \bibinfo {author} {\bibfnamefont {H.~R.}\ \bibnamefont
  {Quinn}},\ }\href {\doibase 10.1103/PhysRevD.16.1791} {\bibfield  {journal}
  {\bibinfo  {journal} {Phys. Rev.}\ }\textbf {\bibinfo {volume} {D16}},\
  \bibinfo {pages} {1791} (\bibinfo {year} {1977}{\natexlab{b}})}\BibitemShut
  {NoStop}%
\bibitem [{\citenamefont {Weinberg}(1978)}]{Weinberg:1977ma}%
  \BibitemOpen
  \bibfield  {author} {\bibinfo {author} {\bibfnamefont {S.}~\bibnamefont
  {Weinberg}},\ }\href {\doibase 10.1103/PhysRevLett.40.223} {\bibfield
  {journal} {\bibinfo  {journal} {Phys. Rev. Lett.}\ }\textbf {\bibinfo
  {volume} {40}},\ \bibinfo {pages} {223} (\bibinfo {year} {1978})}\BibitemShut
  {NoStop}%
\bibitem [{\citenamefont {Wilczek}(1978)}]{Wilczek:1977pj}%
  \BibitemOpen
  \bibfield  {author} {\bibinfo {author} {\bibfnamefont {F.}~\bibnamefont
  {Wilczek}},\ }\href {\doibase 10.1103/PhysRevLett.40.279} {\bibfield
  {journal} {\bibinfo  {journal} {Phys. Rev. Lett.}\ }\textbf {\bibinfo
  {volume} {40}},\ \bibinfo {pages} {279} (\bibinfo {year} {1978})}\BibitemShut
  {NoStop}%
\bibitem [{\citenamefont {Baker}\ \emph {et~al.}(2006)\citenamefont {Baker}
  \emph {et~al.}}]{Baker:2006ts}%
  \BibitemOpen
  \bibfield  {author} {\bibinfo {author} {\bibfnamefont {C.~A.}\ \bibnamefont
  {Baker}} \emph {et~al.},\ }\href {\doibase 10.1103/PhysRevLett.97.131801}
  {\bibfield  {journal} {\bibinfo  {journal} {Phys. Rev. Lett.}\ }\textbf
  {\bibinfo {volume} {97}},\ \bibinfo {pages} {131801} (\bibinfo {year}
  {2006})},\ \Eprint {http://arxiv.org/abs/hep-ex/0602020}
  {arXiv:hep-ex/0602020 [hep-ex]} \BibitemShut {NoStop}%
\bibitem [{\citenamefont {Georgi}(1986)}]{Georgi:1985hf}%
  \BibitemOpen
  \bibfield  {author} {\bibinfo {author} {\bibfnamefont {H.}~\bibnamefont
  {Georgi}},\ }\href {\doibase 10.1016/0550-3213(86)90092-1} {\bibfield
  {journal} {\bibinfo  {journal} {Nucl. Phys.}\ }\textbf {\bibinfo {volume}
  {B266}},\ \bibinfo {pages} {274} (\bibinfo {year} {1986})}\BibitemShut
  {NoStop}%
\bibitem [{\citenamefont {Han}\ \emph {et~al.}(2010)\citenamefont {Han},
  \citenamefont {Lewis},\ and\ \citenamefont {Liu}}]{Han:2010rf}%
  \BibitemOpen
  \bibfield  {author} {\bibinfo {author} {\bibfnamefont {T.}~\bibnamefont
  {Han}}, \bibinfo {author} {\bibfnamefont {I.}~\bibnamefont {Lewis}}, \ and\
  \bibinfo {author} {\bibfnamefont {Z.}~\bibnamefont {Liu}},\ }\href {\doibase
  10.1007/JHEP12(2010)085} {\bibfield  {journal} {\bibinfo  {journal} {JHEP}\
  }\textbf {\bibinfo {volume} {12}},\ \bibinfo {pages} {085} (\bibinfo {year}
  {2010})},\ \Eprint {http://arxiv.org/abs/1010.4309} {arXiv:1010.4309
  [hep-ph]} \BibitemShut {NoStop}%
\bibitem [{\citenamefont {Cohen}\ \emph {et~al.}(1997)\citenamefont {Cohen},
  \citenamefont {Kaplan},\ and\ \citenamefont {Nelson}}]{Cohen:1997rt}%
  \BibitemOpen
  \bibfield  {author} {\bibinfo {author} {\bibfnamefont {A.~G.}\ \bibnamefont
  {Cohen}}, \bibinfo {author} {\bibfnamefont {D.~B.}\ \bibnamefont {Kaplan}}, \
  and\ \bibinfo {author} {\bibfnamefont {A.~E.}\ \bibnamefont {Nelson}},\
  }\href {\doibase 10.1016/S0370-2693(97)00995-7} {\bibfield  {journal}
  {\bibinfo  {journal} {Phys. Lett.}\ }\textbf {\bibinfo {volume} {B412}},\
  \bibinfo {pages} {301} (\bibinfo {year} {1997})},\ \Eprint
  {http://arxiv.org/abs/hep-ph/9706275} {arXiv:hep-ph/9706275 [hep-ph]}
  \BibitemShut {NoStop}%
\bibitem [{\citenamefont {Luty}(1998)}]{Luty:1997fk}%
  \BibitemOpen
  \bibfield  {author} {\bibinfo {author} {\bibfnamefont {M.~A.}\ \bibnamefont
  {Luty}},\ }\href {\doibase 10.1103/PhysRevD.57.1531} {\bibfield  {journal}
  {\bibinfo  {journal} {Phys. Rev.}\ }\textbf {\bibinfo {volume} {D57}},\
  \bibinfo {pages} {1531} (\bibinfo {year} {1998})},\ \Eprint
  {http://arxiv.org/abs/hep-ph/9706235} {arXiv:hep-ph/9706235 [hep-ph]}
  \BibitemShut {NoStop}%
\bibitem [{\citenamefont {Aad}\ \emph {et~al.}(2015)\citenamefont {Aad} \emph
  {et~al.}}]{Aad:2014aqa}%
  \BibitemOpen
  \bibfield  {author} {\bibinfo {author} {\bibfnamefont {G.}~\bibnamefont
  {Aad}} \emph {et~al.} (\bibinfo {collaboration} {ATLAS}),\ }\href {\doibase
  10.1103/PhysRevD.91.052007} {\bibfield  {journal} {\bibinfo  {journal} {Phys.
  Rev.}\ }\textbf {\bibinfo {volume} {D91}},\ \bibinfo {pages} {052007}
  (\bibinfo {year} {2015})},\ \Eprint {http://arxiv.org/abs/1407.1376}
  {arXiv:1407.1376 [hep-ex]} \BibitemShut {NoStop}%
\bibitem [{\citenamefont {Khachatryan}\ \emph {et~al.}(2015)\citenamefont
  {Khachatryan} \emph {et~al.}}]{Khachatryan:2015sja}%
  \BibitemOpen
  \bibfield  {author} {\bibinfo {author} {\bibfnamefont {V.}~\bibnamefont
  {Khachatryan}} \emph {et~al.} (\bibinfo {collaboration} {CMS}),\ }\href
  {\doibase 10.1103/PhysRevD.91.052009} {\bibfield  {journal} {\bibinfo
  {journal} {Phys. Rev.}\ }\textbf {\bibinfo {volume} {D91}},\ \bibinfo {pages}
  {052009} (\bibinfo {year} {2015})},\ \Eprint
  {http://arxiv.org/abs/1501.04198} {arXiv:1501.04198 [hep-ex]} \BibitemShut
  {NoStop}%
\bibitem [{\citenamefont {Goncalves-Netto}\ \emph {et~al.}(2012)\citenamefont
  {Goncalves-Netto}, \citenamefont {Lopez-Val}, \citenamefont {Mawatari},
  \citenamefont {Plehn},\ and\ \citenamefont
  {Wigmore}}]{GoncalvesNetto:2012nt}%
  \BibitemOpen
  \bibfield  {author} {\bibinfo {author} {\bibfnamefont {D.}~\bibnamefont
  {Goncalves-Netto}}, \bibinfo {author} {\bibfnamefont {D.}~\bibnamefont
  {Lopez-Val}}, \bibinfo {author} {\bibfnamefont {K.}~\bibnamefont {Mawatari}},
  \bibinfo {author} {\bibfnamefont {T.}~\bibnamefont {Plehn}}, \ and\ \bibinfo
  {author} {\bibfnamefont {I.}~\bibnamefont {Wigmore}},\ }\href {\doibase
  10.1103/PhysRevD.85.114024} {\bibfield  {journal} {\bibinfo  {journal} {Phys.
  Rev.}\ }\textbf {\bibinfo {volume} {D85}},\ \bibinfo {pages} {114024}
  (\bibinfo {year} {2012})},\ \Eprint {http://arxiv.org/abs/1203.6358}
  {arXiv:1203.6358 [hep-ph]} \BibitemShut {NoStop}%
\bibitem [{\citenamefont {Belyaev}\ \emph {et~al.}(2017)\citenamefont
  {Belyaev}, \citenamefont {Cacciapaglia}, \citenamefont {Cai}, \citenamefont
  {Ferretti}, \citenamefont {Flacke}, \citenamefont {Parolini},\ and\
  \citenamefont {Serodio}}]{Belyaev:2016ftv}%
  \BibitemOpen
  \bibfield  {author} {\bibinfo {author} {\bibfnamefont {A.}~\bibnamefont
  {Belyaev}}, \bibinfo {author} {\bibfnamefont {G.}~\bibnamefont
  {Cacciapaglia}}, \bibinfo {author} {\bibfnamefont {H.}~\bibnamefont {Cai}},
  \bibinfo {author} {\bibfnamefont {G.}~\bibnamefont {Ferretti}}, \bibinfo
  {author} {\bibfnamefont {T.}~\bibnamefont {Flacke}}, \bibinfo {author}
  {\bibfnamefont {A.}~\bibnamefont {Parolini}}, \ and\ \bibinfo {author}
  {\bibfnamefont {H.}~\bibnamefont {Serodio}},\ }\href {\doibase
  10.1007/JHEP01(2017)094} {\bibfield  {journal} {\bibinfo  {journal} {JHEP}\
  }\textbf {\bibinfo {volume} {01}},\ \bibinfo {pages} {094} (\bibinfo {year}
  {2017})},\ \Eprint {http://arxiv.org/abs/1610.06591} {arXiv:1610.06591
  [hep-ph]} \BibitemShut {NoStop}%
\bibitem [{\citenamefont {Khachatryan}\ \emph {et~al.}(2016)\citenamefont
  {Khachatryan} \emph {et~al.}}]{Khachatryan:2016sfv}%
  \BibitemOpen
  \bibfield  {author} {\bibinfo {author} {\bibfnamefont {V.}~\bibnamefont
  {Khachatryan}} \emph {et~al.} (\bibinfo {collaboration} {CMS}),\ }\href
  {\doibase 10.1103/PhysRevD.94.112004} {\bibfield  {journal} {\bibinfo
  {journal} {Phys. Rev.}\ }\textbf {\bibinfo {volume} {D94}},\ \bibinfo {pages}
  {112004} (\bibinfo {year} {2016})},\ \Eprint
  {http://arxiv.org/abs/1609.08382} {arXiv:1609.08382 [hep-ex]} \BibitemShut
  {NoStop}%
\bibitem [{\citenamefont {Bizot}\ and\ \citenamefont
  {Frigerio}(2016)}]{Bizot:2015zaa}%
  \BibitemOpen
  \bibfield  {author} {\bibinfo {author} {\bibfnamefont {N.}~\bibnamefont
  {Bizot}}\ and\ \bibinfo {author} {\bibfnamefont {M.}~\bibnamefont
  {Frigerio}},\ }\href {\doibase 10.1007/JHEP01(2016)036} {\bibfield  {journal}
  {\bibinfo  {journal} {JHEP}\ }\textbf {\bibinfo {volume} {01}},\ \bibinfo
  {pages} {036} (\bibinfo {year} {2016})},\ \Eprint
  {http://arxiv.org/abs/1508.01645} {arXiv:1508.01645 [hep-ph]} \BibitemShut
  {NoStop}%
\bibitem [{ATL(2015)}]{ATLAS-CONF-2015-012}%
  \BibitemOpen
  \href {http://cds.cern.ch/record/2002556} {\emph {\bibinfo {title} {{Search
  for production of vector-like quark pairs and of four top quarks in the
  lepton plus jets final state in $pp$ collisions at $\sqrt{s}=8$ TeV with the
  ATLAS detector}}}},\ \bibinfo {type} {Tech. Rep.}\ \bibinfo {number}
  {ATLAS-CONF-2015-012}\ (\bibinfo  {institution} {CERN},\ \bibinfo {address}
  {Geneva},\ \bibinfo {year} {2015})\BibitemShut {NoStop}%
\bibitem [{ATL(2016)}]{ATLAS-CONF-2016-013}%
  \BibitemOpen
  \href {http://cds.cern.ch/record/2140998} {\emph {\bibinfo {title} {{Search
  for production of vector-like top quark pairs and of four top quarks in the
  lepton-plus-jets final state in $pp$ collisions at $\sqrt{s}=13$ TeV with the
  ATLAS detector}}}},\ \bibinfo {type} {Tech. Rep.}\ \bibinfo {number}
  {ATLAS-CONF-2016-013}\ (\bibinfo  {institution} {CERN},\ \bibinfo {address}
  {Geneva},\ \bibinfo {year} {2016})\BibitemShut {NoStop}%
\bibitem [{CMS(2013)}]{CMS-PAS-B2G-12-015}%
  \BibitemOpen
  \href {http://cds.cern.ch/record/1557571} {\emph {\bibinfo {title}
  {{Inclusive search for a vector-like T quark by CMS}}}},\ \bibinfo {type}
  {Tech. Rep.}\ \bibinfo {number} {CMS-PAS-B2G-12-015}\ (\bibinfo
  {institution} {CERN},\ \bibinfo {address} {Geneva},\ \bibinfo {year}
  {2013})\BibitemShut {NoStop}%
\bibitem [{CMS(2016)}]{CMS-PAS-B2G-16-002}%
  \BibitemOpen
  \href {https://cds.cern.ch/record/2141070} {\emph {\bibinfo {title} {{Search
  for pair production of vector-like T quarks in the lepton plus jets final
  state}}}},\ \bibinfo {type} {Tech. Rep.}\ \bibinfo {number}
  {CMS-PAS-B2G-16-002}\ (\bibinfo  {institution} {CERN},\ \bibinfo {address}
  {Geneva},\ \bibinfo {year} {2016})\BibitemShut {NoStop}%
\bibitem [{\citenamefont {Collaboration}(2013)}]{CMS:2013yfa}%
  \BibitemOpen
  \bibfield  {author} {\bibinfo {author} {\bibfnamefont {C.}~\bibnamefont
  {Collaboration}} (\bibinfo {collaboration} {CMS}),\ }\href@noop {} {\
  (\bibinfo {year} {2013})}\BibitemShut {NoStop}%
\bibitem [{\citenamefont {Sikivie}(2008)}]{Sikivie:2006ni}%
  \BibitemOpen
  \bibfield  {author} {\bibinfo {author} {\bibfnamefont {P.}~\bibnamefont
  {Sikivie}},\ }\bibfield  {booktitle} {\emph {\bibinfo {booktitle} {{Axions:
  Theory, cosmology, and experimental searches. Proceedings, 1st Joint
  ILIAS-CERN-CAST axion training, Geneva, Switzerland, November 30-December 2,
  2005}}},\ }\href {\doibase 10.1007/978-3-540-73518-2_2} {\bibfield  {journal}
  {\bibinfo  {journal} {Lect. Notes Phys.}\ }\textbf {\bibinfo {volume}
  {741}},\ \bibinfo {pages} {19} (\bibinfo {year} {2008})},\ \bibinfo {note}
  {[,19(2006)]},\ \Eprint {http://arxiv.org/abs/astro-ph/0610440}
  {arXiv:astro-ph/0610440 [astro-ph]} \BibitemShut {NoStop}%
\bibitem [{\citenamefont {Ade}\ \emph {et~al.}(2016)\citenamefont {Ade} \emph
  {et~al.}}]{Ade:2015xua}%
  \BibitemOpen
  \bibfield  {author} {\bibinfo {author} {\bibfnamefont {P.~A.~R.}\
  \bibnamefont {Ade}} \emph {et~al.} (\bibinfo {collaboration} {Planck}),\
  }\href {\doibase 10.1051/0004-6361/201525830} {\bibfield  {journal} {\bibinfo
   {journal} {Astron. Astrophys.}\ }\textbf {\bibinfo {volume} {594}},\
  \bibinfo {pages} {A13} (\bibinfo {year} {2016})},\ \Eprint
  {http://arxiv.org/abs/1502.01589} {arXiv:1502.01589 [astro-ph.CO]}
  \BibitemShut {NoStop}%
\bibitem [{\citenamefont {Sikivie}(1983)}]{Sikivie:1983ip}%
  \BibitemOpen
  \bibfield  {author} {\bibinfo {author} {\bibfnamefont {P.}~\bibnamefont
  {Sikivie}},\ }\bibfield  {booktitle} {\emph {\bibinfo {booktitle} {{11th
  International Symposium on Lepton and Photon Interactions at High Energies
  Ithaca, New York, August 4-9, 1983}}},\ }\href {\doibase
  10.1103/PhysRevLett.51.1415, 10.1103/PhysRevLett.52.695.2} {\bibfield
  {journal} {\bibinfo  {journal} {Phys. Rev. Lett.}\ }\textbf {\bibinfo
  {volume} {51}},\ \bibinfo {pages} {1415} (\bibinfo {year} {1983})},\ \bibinfo
  {note} {[Erratum: Phys. Rev. Lett.52,695(1984)]}\BibitemShut {NoStop}%
\bibitem [{\citenamefont {Bradley}\ \emph {et~al.}(2003)\citenamefont
  {Bradley}, \citenamefont {Clarke}, \citenamefont {Kinion}, \citenamefont
  {Rosenberg}, \citenamefont {van Bibber}, \citenamefont {Matsuki},
  \citenamefont {Muck},\ and\ \citenamefont {Sikivie}}]{Bradley:2003kg}%
  \BibitemOpen
  \bibfield  {author} {\bibinfo {author} {\bibfnamefont {R.}~\bibnamefont
  {Bradley}}, \bibinfo {author} {\bibfnamefont {J.}~\bibnamefont {Clarke}},
  \bibinfo {author} {\bibfnamefont {D.}~\bibnamefont {Kinion}}, \bibinfo
  {author} {\bibfnamefont {L.~J.}\ \bibnamefont {Rosenberg}}, \bibinfo {author}
  {\bibfnamefont {K.}~\bibnamefont {van Bibber}}, \bibinfo {author}
  {\bibfnamefont {S.}~\bibnamefont {Matsuki}}, \bibinfo {author} {\bibfnamefont
  {M.}~\bibnamefont {Muck}}, \ and\ \bibinfo {author} {\bibfnamefont
  {P.}~\bibnamefont {Sikivie}},\ }\href {\doibase 10.1103/RevModPhys.75.777}
  {\bibfield  {journal} {\bibinfo  {journal} {Rev. Mod. Phys.}\ }\textbf
  {\bibinfo {volume} {75}},\ \bibinfo {pages} {777} (\bibinfo {year}
  {2003})}\BibitemShut {NoStop}%
\bibitem [{\citenamefont {Asztalos}\ \emph {et~al.}(2010)\citenamefont
  {Asztalos} \emph {et~al.}}]{Asztalos:2009yp}%
  \BibitemOpen
  \bibfield  {author} {\bibinfo {author} {\bibfnamefont {S.~J.}\ \bibnamefont
  {Asztalos}} \emph {et~al.} (\bibinfo {collaboration} {ADMX}),\ }\href
  {\doibase 10.1103/PhysRevLett.104.041301} {\bibfield  {journal} {\bibinfo
  {journal} {Phys. Rev. Lett.}\ }\textbf {\bibinfo {volume} {104}},\ \bibinfo
  {pages} {041301} (\bibinfo {year} {2010})},\ \Eprint
  {http://arxiv.org/abs/0910.5914} {arXiv:0910.5914 [astro-ph.CO]} \BibitemShut
  {NoStop}%
\bibitem [{\citenamefont {Stern}(2016)}]{Stern:2016bbw}%
  \BibitemOpen
  \bibfield  {author} {\bibinfo {author} {\bibfnamefont {I.}~\bibnamefont
  {Stern}},\ }\bibfield  {booktitle} {\emph {\bibinfo {booktitle}
  {{Proceedings, 38th International Conference on High Energy Physics (ICHEP
  2016): Chicago, IL, USA, August 3-10, 2016}}},\ }\href@noop {} {\bibfield
  {journal} {\bibinfo  {journal} {PoS}\ }\textbf {\bibinfo {volume}
  {ICHEP2016}},\ \bibinfo {pages} {198} (\bibinfo {year} {2016})},\ \Eprint
  {http://arxiv.org/abs/1612.08296} {arXiv:1612.08296 [physics.ins-det]}
  \BibitemShut {NoStop}%
\bibitem [{\citenamefont {Gelmini}\ \emph {et~al.}(1989)\citenamefont
  {Gelmini}, \citenamefont {Gleiser},\ and\ \citenamefont
  {Kolb}}]{Gelmini:1988sf}%
  \BibitemOpen
  \bibfield  {author} {\bibinfo {author} {\bibfnamefont {G.~B.}\ \bibnamefont
  {Gelmini}}, \bibinfo {author} {\bibfnamefont {M.}~\bibnamefont {Gleiser}}, \
  and\ \bibinfo {author} {\bibfnamefont {E.~W.}\ \bibnamefont {Kolb}},\ }\href
  {\doibase 10.1103/PhysRevD.39.1558} {\bibfield  {journal} {\bibinfo
  {journal} {Phys. Rev.}\ }\textbf {\bibinfo {volume} {D39}},\ \bibinfo {pages}
  {1558} (\bibinfo {year} {1989})}\BibitemShut {NoStop}%
\bibitem [{\citenamefont {Larsson}\ \emph {et~al.}(1997)\citenamefont
  {Larsson}, \citenamefont {Sarkar},\ and\ \citenamefont
  {White}}]{Larsson:1996sp}%
  \BibitemOpen
  \bibfield  {author} {\bibinfo {author} {\bibfnamefont {S.~E.}\ \bibnamefont
  {Larsson}}, \bibinfo {author} {\bibfnamefont {S.}~\bibnamefont {Sarkar}}, \
  and\ \bibinfo {author} {\bibfnamefont {P.~L.}\ \bibnamefont {White}},\ }\href
  {\doibase 10.1103/PhysRevD.55.5129} {\bibfield  {journal} {\bibinfo
  {journal} {Phys. Rev.}\ }\textbf {\bibinfo {volume} {D55}},\ \bibinfo {pages}
  {5129} (\bibinfo {year} {1997})},\ \Eprint
  {http://arxiv.org/abs/hep-ph/9608319} {arXiv:hep-ph/9608319 [hep-ph]}
  \BibitemShut {NoStop}%
\bibitem [{\citenamefont {Hiramatsu}\ \emph {et~al.}(2013)\citenamefont
  {Hiramatsu}, \citenamefont {Kawasaki}, \citenamefont {Saikawa},\ and\
  \citenamefont {Sekiguchi}}]{Hiramatsu:2012sc}%
  \BibitemOpen
  \bibfield  {author} {\bibinfo {author} {\bibfnamefont {T.}~\bibnamefont
  {Hiramatsu}}, \bibinfo {author} {\bibfnamefont {M.}~\bibnamefont {Kawasaki}},
  \bibinfo {author} {\bibfnamefont {K.}~\bibnamefont {Saikawa}}, \ and\
  \bibinfo {author} {\bibfnamefont {T.}~\bibnamefont {Sekiguchi}},\ }\href
  {\doibase 10.1088/1475-7516/2013/01/001} {\bibfield  {journal} {\bibinfo
  {journal} {JCAP}\ }\textbf {\bibinfo {volume} {1301}},\ \bibinfo {pages}
  {001} (\bibinfo {year} {2013})},\ \Eprint {http://arxiv.org/abs/1207.3166}
  {arXiv:1207.3166 [hep-ph]} \BibitemShut {NoStop}%
\bibitem [{\citenamefont {Griest}\ and\ \citenamefont
  {Kamionkowski}(1990)}]{Griest:1989wd}%
  \BibitemOpen
  \bibfield  {author} {\bibinfo {author} {\bibfnamefont {K.}~\bibnamefont
  {Griest}}\ and\ \bibinfo {author} {\bibfnamefont {M.}~\bibnamefont
  {Kamionkowski}},\ }\href {\doibase 10.1103/PhysRevLett.64.615} {\bibfield
  {journal} {\bibinfo  {journal} {Phys. Rev. Lett.}\ }\textbf {\bibinfo
  {volume} {64}},\ \bibinfo {pages} {615} (\bibinfo {year} {1990})}\BibitemShut
  {NoStop}%
\bibitem [{\citenamefont {Huo}\ \emph {et~al.}(2016)\citenamefont {Huo},
  \citenamefont {Matsumoto}, \citenamefont {Sming~Tsai},\ and\ \citenamefont
  {Yanagida}}]{Huo:2015nwa}%
  \BibitemOpen
  \bibfield  {author} {\bibinfo {author} {\bibfnamefont {R.}~\bibnamefont
  {Huo}}, \bibinfo {author} {\bibfnamefont {S.}~\bibnamefont {Matsumoto}},
  \bibinfo {author} {\bibfnamefont {Y.-L.}\ \bibnamefont {Sming~Tsai}}, \ and\
  \bibinfo {author} {\bibfnamefont {T.~T.}\ \bibnamefont {Yanagida}},\ }\href
  {\doibase 10.1007/JHEP09(2016)162} {\bibfield  {journal} {\bibinfo  {journal}
  {JHEP}\ }\textbf {\bibinfo {volume} {09}},\ \bibinfo {pages} {162} (\bibinfo
  {year} {2016})},\ \Eprint {http://arxiv.org/abs/1506.06929} {arXiv:1506.06929
  [hep-ph]} \BibitemShut {NoStop}%
\bibitem [{\citenamefont {Freese}\ \emph {et~al.}(1990)\citenamefont {Freese},
  \citenamefont {Frieman},\ and\ \citenamefont {Olinto}}]{Freese:1990rb}%
  \BibitemOpen
  \bibfield  {author} {\bibinfo {author} {\bibfnamefont {K.}~\bibnamefont
  {Freese}}, \bibinfo {author} {\bibfnamefont {J.~A.}\ \bibnamefont {Frieman}},
  \ and\ \bibinfo {author} {\bibfnamefont {A.~V.}\ \bibnamefont {Olinto}},\
  }\href {\doibase 10.1103/PhysRevLett.65.3233} {\bibfield  {journal} {\bibinfo
   {journal} {Phys. Rev. Lett.}\ }\textbf {\bibinfo {volume} {65}},\ \bibinfo
  {pages} {3233} (\bibinfo {year} {1990})}\BibitemShut {NoStop}%
\bibitem [{\citenamefont {{Neronov}}\ and\ \citenamefont
  {{Vovk}}(2010)}]{2010Sci...328...73N}%
  \BibitemOpen
  \bibfield  {author} {\bibinfo {author} {\bibfnamefont {A.}~\bibnamefont
  {{Neronov}}}\ and\ \bibinfo {author} {\bibfnamefont {I.}~\bibnamefont
  {{Vovk}}},\ }\href {\doibase 10.1126/science.1184192} {\bibfield  {journal}
  {\bibinfo  {journal} {Science}\ }\textbf {\bibinfo {volume} {328}},\ \bibinfo
  {pages} {73} (\bibinfo {year} {2010})},\ \Eprint
  {http://arxiv.org/abs/1006.3504} {arXiv:1006.3504 [astro-ph.HE]} \BibitemShut
  {NoStop}%
\bibitem [{\citenamefont {{Tavecchio}}\ \emph {et~al.}()\citenamefont
  {{Tavecchio}}, \citenamefont {{Ghisellini}}, \citenamefont {{Foschini}},
  \citenamefont {{Bonnoli}}, \citenamefont {{Ghirlanda}},\ and\ \citenamefont
  {{Coppi}}}]{2010MNRAS.406L..70T}%
  \BibitemOpen
  \bibfield  {author} {\bibinfo {author} {\bibfnamefont {F.}~\bibnamefont
  {{Tavecchio}}}, \bibinfo {author} {\bibfnamefont {G.}~\bibnamefont
  {{Ghisellini}}}, \bibinfo {author} {\bibfnamefont {L.}~\bibnamefont
  {{Foschini}}}, \bibinfo {author} {\bibfnamefont {G.}~\bibnamefont
  {{Bonnoli}}}, \bibinfo {author} {\bibfnamefont {G.}~\bibnamefont
  {{Ghirlanda}}}, \ and\ \bibinfo {author} {\bibfnamefont {P.}~\bibnamefont
  {{Coppi}}},\ }\href@noop {} {\ }\BibitemShut {NoStop}%
\bibitem [{\citenamefont {{Dolag}}\ \emph {et~al.}()\citenamefont {{Dolag}},
  \citenamefont {{Kachelriess}}, \citenamefont {{Ostapchenko}},\ and\
  \citenamefont {{Tom{\`a}s}}}]{2011ApJ...727L...4D}%
  \BibitemOpen
  \bibfield  {author} {\bibinfo {author} {\bibfnamefont {K.}~\bibnamefont
  {{Dolag}}}, \bibinfo {author} {\bibfnamefont {M.}~\bibnamefont
  {{Kachelriess}}}, \bibinfo {author} {\bibfnamefont {S.}~\bibnamefont
  {{Ostapchenko}}}, \ and\ \bibinfo {author} {\bibfnamefont {R.}~\bibnamefont
  {{Tom{\`a}s}}},\ }\href@noop {} {\ }\BibitemShut {NoStop}%
\bibitem [{\citenamefont {Taylor}\ \emph {et~al.}(2011)\citenamefont {Taylor},
  \citenamefont {Vovk},\ and\ \citenamefont {Neronov}}]{Taylor:2011bn}%
  \BibitemOpen
  \bibfield  {author} {\bibinfo {author} {\bibfnamefont {A.~M.}\ \bibnamefont
  {Taylor}}, \bibinfo {author} {\bibfnamefont {I.}~\bibnamefont {Vovk}}, \ and\
  \bibinfo {author} {\bibfnamefont {A.}~\bibnamefont {Neronov}},\ }\href
  {\doibase 10.1051/0004-6361/201116441} {\bibfield  {journal} {\bibinfo
  {journal} {Astron. Astrophys.}\ }\textbf {\bibinfo {volume} {529}},\ \bibinfo
  {pages} {A144} (\bibinfo {year} {2011})},\ \Eprint
  {http://arxiv.org/abs/1101.0932} {arXiv:1101.0932 [astro-ph.HE]} \BibitemShut
  {NoStop}%
\bibitem [{\citenamefont {Essey}\ \emph {et~al.}(2011)\citenamefont {Essey},
  \citenamefont {Ando},\ and\ \citenamefont {Kusenko}}]{Essey:2010nd}%
  \BibitemOpen
  \bibfield  {author} {\bibinfo {author} {\bibfnamefont {W.}~\bibnamefont
  {Essey}}, \bibinfo {author} {\bibfnamefont {S.}~\bibnamefont {Ando}}, \ and\
  \bibinfo {author} {\bibfnamefont {A.}~\bibnamefont {Kusenko}},\ }\href
  {\doibase 10.1016/j.astropartphys.2011.06.010} {\bibfield  {journal}
  {\bibinfo  {journal} {Astropart. Phys.}\ }\textbf {\bibinfo {volume} {35}},\
  \bibinfo {pages} {135} (\bibinfo {year} {2011})},\ \Eprint
  {http://arxiv.org/abs/1012.5313} {arXiv:1012.5313 [astro-ph.HE]} \BibitemShut
  {NoStop}%
\bibitem [{\citenamefont {Takahashi}\ \emph {et~al.}(2013)\citenamefont
  {Takahashi}, \citenamefont {Mori}, \citenamefont {Ichiki}, \citenamefont
  {Inoue},\ and\ \citenamefont {Takami}}]{Takahashi:2013lba}%
  \BibitemOpen
  \bibfield  {author} {\bibinfo {author} {\bibfnamefont {K.}~\bibnamefont
  {Takahashi}}, \bibinfo {author} {\bibfnamefont {M.}~\bibnamefont {Mori}},
  \bibinfo {author} {\bibfnamefont {K.}~\bibnamefont {Ichiki}}, \bibinfo
  {author} {\bibfnamefont {S.}~\bibnamefont {Inoue}}, \ and\ \bibinfo {author}
  {\bibfnamefont {H.}~\bibnamefont {Takami}},\ }\href {\doibase
  10.1088/2041-8205/771/2/L42} {\bibfield  {journal} {\bibinfo  {journal}
  {Astrophys. J.}\ }\textbf {\bibinfo {volume} {771}},\ \bibinfo {pages} {L42}
  (\bibinfo {year} {2013})},\ \Eprint {http://arxiv.org/abs/1303.3069}
  {arXiv:1303.3069 [astro-ph.CO]} \BibitemShut {NoStop}%
\bibitem [{\citenamefont {Finke}\ \emph {et~al.}(2012)\citenamefont {Finke},
  \citenamefont {Reyes},\ and\ \citenamefont {Georganopoulos}}]{Finke:2013tyq}%
  \BibitemOpen
  \bibfield  {author} {\bibinfo {author} {\bibfnamefont {J.}~\bibnamefont
  {Finke}}, \bibinfo {author} {\bibfnamefont {L.}~\bibnamefont {Reyes}}, \ and\
  \bibinfo {author} {\bibfnamefont {M.}~\bibnamefont {Georganopoulos}}
  (\bibinfo {collaboration} {Fermi-LAT}),\ }\bibfield  {booktitle} {\emph
  {\bibinfo {booktitle} {{Proceedings, 4th International Fermi Symposium:
  Monterey, California, USA, October 28-November 2, 2012}}},\ }\href@noop {}
  {\bibfield  {journal} {\bibinfo  {journal} {eConf}\ }\textbf {\bibinfo
  {volume} {C121028}},\ \bibinfo {pages} {365} (\bibinfo {year} {2012})},\
  \Eprint {http://arxiv.org/abs/1303.5093} {arXiv:1303.5093 [astro-ph.HE]}
  \BibitemShut {NoStop}%
\bibitem [{\citenamefont {Finke}\ \emph {et~al.}(2015)\citenamefont {Finke},
  \citenamefont {Reyes}, \citenamefont {Georganopoulos}, \citenamefont
  {Reynolds}, \citenamefont {Ajello}, \citenamefont {Fegan},\ and\
  \citenamefont {McCann}}]{Finke:2015ona}%
  \BibitemOpen
  \bibfield  {author} {\bibinfo {author} {\bibfnamefont {J.~D.}\ \bibnamefont
  {Finke}}, \bibinfo {author} {\bibfnamefont {L.~C.}\ \bibnamefont {Reyes}},
  \bibinfo {author} {\bibfnamefont {M.}~\bibnamefont {Georganopoulos}},
  \bibinfo {author} {\bibfnamefont {K.}~\bibnamefont {Reynolds}}, \bibinfo
  {author} {\bibfnamefont {M.}~\bibnamefont {Ajello}}, \bibinfo {author}
  {\bibfnamefont {S.~J.}\ \bibnamefont {Fegan}}, \ and\ \bibinfo {author}
  {\bibfnamefont {K.}~\bibnamefont {McCann}},\ }\href {\doibase
  10.1088/0004-637X/814/1/20} {\bibfield  {journal} {\bibinfo  {journal}
  {Astrophys. J.}\ }\textbf {\bibinfo {volume} {814}},\ \bibinfo {pages} {20}
  (\bibinfo {year} {2015})},\ \Eprint {http://arxiv.org/abs/1510.02485}
  {arXiv:1510.02485 [astro-ph.HE]} \BibitemShut {NoStop}%
\bibitem [{\citenamefont {Fujita}\ \emph {et~al.}(2015)\citenamefont {Fujita},
  \citenamefont {Namba}, \citenamefont {Tada}, \citenamefont {Takeda},\ and\
  \citenamefont {Tashiro}}]{Fujita:2015iga}%
  \BibitemOpen
  \bibfield  {author} {\bibinfo {author} {\bibfnamefont {T.}~\bibnamefont
  {Fujita}}, \bibinfo {author} {\bibfnamefont {R.}~\bibnamefont {Namba}},
  \bibinfo {author} {\bibfnamefont {Y.}~\bibnamefont {Tada}}, \bibinfo {author}
  {\bibfnamefont {N.}~\bibnamefont {Takeda}}, \ and\ \bibinfo {author}
  {\bibfnamefont {H.}~\bibnamefont {Tashiro}},\ }\href {\doibase
  10.1088/1475-7516/2015/05/054} {\bibfield  {journal} {\bibinfo  {journal}
  {JCAP}\ }\textbf {\bibinfo {volume} {1505}},\ \bibinfo {pages} {054}
  (\bibinfo {year} {2015})},\ \Eprint {http://arxiv.org/abs/1503.05802}
  {arXiv:1503.05802 [astro-ph.CO]} \BibitemShut {NoStop}%
\bibitem [{\citenamefont {Adshead}\ \emph {et~al.}(2016)\citenamefont
  {Adshead}, \citenamefont {Giblin}, \citenamefont {Scully},\ and\
  \citenamefont {Sfakianakis}}]{Adshead:2016iae}%
  \BibitemOpen
  \bibfield  {author} {\bibinfo {author} {\bibfnamefont {P.}~\bibnamefont
  {Adshead}}, \bibinfo {author} {\bibfnamefont {J.~T.}\ \bibnamefont {Giblin}},
  \bibinfo {author} {\bibfnamefont {T.~R.}\ \bibnamefont {Scully}}, \ and\
  \bibinfo {author} {\bibfnamefont {E.~I.}\ \bibnamefont {Sfakianakis}},\
  }\href {\doibase 10.1088/1475-7516/2016/10/039} {\bibfield  {journal}
  {\bibinfo  {journal} {JCAP}\ }\textbf {\bibinfo {volume} {1610}},\ \bibinfo
  {pages} {039} (\bibinfo {year} {2016})},\ \Eprint
  {http://arxiv.org/abs/1606.08474} {arXiv:1606.08474 [astro-ph.CO]}
  \BibitemShut {NoStop}%
\end{thebibliography}%
\end{document}